\documentclass[floatfix,aip,amsmath,amssymb,preprint]{revtex4-1}

\usepackage{graphicx}%
\usepackage{dcolumn}%
\usepackage{bm}%
\usepackage{mathtools}
\usepackage[utf8]{inputenc}

\newcommand{\minus}{\scalebox{0.75}[1.0]{$-$}}

\usepackage [english]{babel}
\usepackage [autostyle, english = american]{csquotes}

\usepackage{calrsfs}
\DeclareMathAlphabet{\pazocal}{OMS}{zplm}{m}{n}
\SetMathAlphabet\pazocal{bold}{OMS}{zplm}{bx}{n}

\newcommand{\hide}[1]{\ignorespaces}

\usepackage{xcolor}
\newcommand\todo[1]{\textcolor{red}{#1}}

\usepackage[normalem]{ulem}

\usepackage[T1]{fontenc}
\usepackage{mathptmx}
\usepackage{etoolbox}
\usepackage{tikz}

\makeatletter
\def\@email#1#2{%
 \endgroup
 \patchcmd{\titleblock@produce}
  {\frontmatter@RRAPformat}
  {\frontmatter@RRAPformat{\produce@RRAP{*#1\href{mailto:#2}{#2}}}\frontmatter@RRAPformat}
  {}{}
}%
\makeatother

\usepackage{fancyhdr}
\begin{document}

\preprint{AIP/123-QED}

\title[]
{Approaching Periodic Systems in Ensemble Density Functional Theory \textit{via} Finite One-Dimensional Models}
\author{Remi J. Leano}

\author{Aurora Pribram-Jones}%
\affiliation{ 
Department of Chemistry and Biochemistry, University of California, Merced, 5200 N. Lake Rd., Merced, CA 95343, USA%
}%
\email{apj@ucmerced.edu}

\author{David A. Strubbe}
\affiliation{%
Department of Physics, University of California, Merced, 5200 N. Lake Rd., Merced, CA 95343, USA %
}%
\email{dstrubbe@ucmerced.edu}

\date{\today}%

\begin{abstract}
Ensemble Density Functional Theory (EDFT) is a generalization of ground-state Density Functional Theory (GS DFT), which is based on an exact formal theory of finite collections of a system's ground and excited states. EDFT in various forms has been shown to improve the accuracy of calculated energy level differences in isolated model systems, atoms, and molecules, but it is not yet clear how EDFT could be used to calculate band gaps for periodic systems. We extend the application of EDFT toward periodic systems by estimating the thermodynamic limit with increasingly large finite one-dimensional 
\enquote{particle in a box} systems, which approach the uniform electron gas (UEG). Using ensemble-generalized Hartree and Local Spin Density Approximation (LSDA) exchange-correlation functionals, we find that corrections go to zero in the infinite limit, as expected for a metallic system.
However, there is a correction to the effective mass, with results comparable to other calculations on 1D, 2D, and 3D UEGs, which indicates promise for non-trivial results from EDFT on periodic systems.

\end{abstract}

\maketitle

\section{\label{sec:Introduction}Introduction}

A well known difficulty with ground-state (GS) 
Density Functional Theory (DFT) is the band gap problem, where the difference between the highest occupied \hide{(HO)} and lowest unoccupied \hide{(LU)} Kohn-Sham (KS) energy states is smaller than the true band gap.\cite{SF18, PL83} There are several methods used to extend GS DFT to excited states, including time-dependent DFT (TDDFT)\cite{RG84, U12, MMNGR12} and the $\Delta$SCF method.\cite{10.1002/qua.560100803,10.1080/00268977400102581,10.1002/qua.560210123} TDDFT has become the standard method for calculating the excitation energies of molecules, achieving accuracies comparable to quantities in GS DFT.\cite{JPC10, C96, HIRC11, EGCM11} However, in its typical application within the adiabatic approximation, TDDFT inadequately describes double and multiple excitations,\cite{doi:10.1146} and struggles with periodic systems.  Typical approximations to the exchange-correlation (XC) kernel $f_{\rm xc}$ lack the correct long-range behavior, which indeed goes to zero in the local-density approximation (LDA).\cite{ORR02, MMNGR12, U12, UY14, MZCB04, DWH03, Botti_2007}
 Similarly, the correction to excitation energies provided by the $\Delta$SCF method for standard XC approximations goes to zero in periodic systems,\cite{PhysRevLett.51.1888, GW98_PRL} which some methods have been proposed to solve.\cite{CG10} The theory of ensemble DFT (EDFT) is another DFT approach to excited states which could be promising for periodic systems, but it remains to be seen how the theory as formulated by Theophilou\cite{MD87} and later by Gross, Oliveira, and Kohn\cite{GOK88I} can be properly formulated for such systems.

Like GS DFT, EDFT is based on a variational theorem. The difference in the two theories is that while in GS DFT, the GS energy is a functional of the GS density, in EDFT, the ensemble energy is a functional of both the ensemble density and a set of ensemble weights, providing access to excited-state quantities.\cite{GOK88_2805,GOK88I,SB18-b} Excitation energies can, in theory, be extracted from the total ensemble energy, and EDFT can account for the discontinuous nature of the XC potential through explicit dependence on weights.\cite{DMF17, SF18, KK13, PPLB82, cernatic2024neutral} Thus EDFT offers a non-perturbative alternative to TDDFT which can more easily treat multiple- and charge-transfer excitations. \hide{Because EDFT is based on the thermodynamic ensemble, it allows for the treatment of both open systems (variable particle number) and excited states.}  Additionally, EDFT can treat both the fundamental (charged)\cite{KK13, SF18} and optical (neutral)\cite{GOK88I} gaps of systems. Relatively accurate EDFT calculations have been performed for small atoms,\cite{YTPB14, YPB17} the hydrogen molecule, \cite{BTH15} for two electrons in boxes or in a 3D harmonic well (Hooke's atom),\cite{PYTB14} the asymmetric Hubbard dimer,\cite{DMSF18,DF19,DMF17} and for some molecules.\cite{Filatov15,doi:10.1021/acs.jpclett.2c00042} However, developing the necessary weight-dependent functionals in order to use EDFT is a complicated task that remains at an early stage of development and limits EDFT's application to a wider span of systems.\cite{GOK88II, PhysRevA.34.737,N98, GPG02,https://doi.org/10.1002/qua.560560833,MSFL20,PhysRevA.104.052806, GSP20, PhysRevLett.123.016401} The key difficulty of EDFT for periodic systems is that the excited states of solids are a continuum of states and as such cannot be modelled with existing EDFT approaches which construct the ensemble from a finite number of individual states.\cite{GOK88I} 

In this paper, rather than studying a periodic system in EDFT directly, we study EDFT by means of finite one-dimensional (1D) systems approaching the thermodynamic limit, performing DFT calculations in the open-source real-space code Octopus.\cite{Octopus_PCCP,Tea20} In section \ref{sec:1Dwell} we introduce a 1D system whose KS potential is a \enquote{particle in a box} (PIB). We build ensembles for the system with the %
weighting scheme described in section \ref{subsec:EDFT}. We motivate our choice of weight-dependent functionals in section \ref{subsec:HXC} and describe the multiplet structure and construction of many-electron densities for our system in section \ref{subsec:Slater}. In sections \ref{sec:UEG} and \ref{subsec:Limit} we outline the necessity of studying systems in the approach to the thermodynamic limit rather than direct study of periodic systems within EDFT. In section \ref{sec:Methods}, we describe our computational methodology for calculations of first (triplet) and second (singlet) excitation energies. Finally, in section \ref{sec:Results}, we discuss ensemble corrections to excitation energies and effective masses obtained in the approach to the thermodynamic limit. We find non-trivial renormalization of the effective masses with results from the tri-ensemble similar to the uniform electron gas in other dimensionalities, showing the promise of EDFT for describing periodic systems.

\section{\label{sec:Theory}Theory}

\subsection{\label{sec:1Dwell}The 1D PIB Potential is the KS Potential}

The PIB potential is defined as a free particle within the confines of a box of length $2L$, subject to an infinite potential outside these boundaries:
\begin{equation}
V(x) =
    \begin{cases}
            0, \hspace{0.5em} -L < x < L, \\
            \infty, \hspace{0.5em} x \leq -L \hspace{0.5em} \text{or} \hspace{0.5em} x \geq L.
    \end{cases}
    \label{eqn:PIBpot}
\end{equation}
The PIB is more readily adaptable to study in the thermodynamic limit than atom-based models, %
and in the limit it becomes the uniform electron gas (UEG) which is a prototypical model in electronic structure theory and is used as a simplified model for the behavior of electrons in metals.\cite{Kaxiras} The 1D UEG is known from the Lieb-Mattis theorem \cite{Lieb-Mattis} to be a singlet at all densities, which has also been found in quantum Monte Carlo calculations. \cite{Malatesta,Malatesta_thesis} It is also expected to be metallic according to Luttinger liquid theory.\cite{Haldane_1981,Imambekov} In this work, we set the KS potential, $v_{\text{KS}}$,
equal to the PIB potential such that $v_{\text{KS}}(x,\sigma) = 0$ within the boundaries of the box. Setting $v_{\text{KS}}$ rather than $v_{\text{ext}}$ to the PIB potential allows us to determine the KS wavefunctions and eigenvalues exactly, and bypasses the need to solve for them self-consistently. A similar approach has been used in studies of a model atom whose KS potential is $1/r$.\cite{Savin1}
In the thermodynamic limit, we obtain the UEG, whether we set $v_{\text{ext}}$ or $v_{\text{KS}}$ equal to the PIB potential. In this limit, the density is constant, leading to a constant $v_{\text{Hxc}} \left[ \rho \right]$, which provides only an overall offset to the eigenvalues and no difference in the excitation energies.

Here we first discuss such a set-up in the context of GS DFT, and then describe the construction of the ensemble in section \ref{subsec:EDFT}. In GS-DFT, the KS potential is defined as
\begin{equation}
    v_{\text{KS}}[\rho](x,\sigma) = v_{\text{ext}}(x) + e^2 \int \frac{ \sum_{\sigma'=\alpha}^\beta \rho(x',\sigma')}{|x-x'|}dx' + \frac{\delta E_{\text{xc}}[\rho]}{\delta \rho(x,\sigma)},
    \label{eqn:vs}
\end{equation}
where $\sigma$ is the spin variable. %
For simplicity, we limit our study to 1D, though a similar procedure could be followed for 2D or 3D. Here the first term on the right is the external potential, the second term is the Hartree potential (where $e$ is the electron charge), and the third term is the XC potential. The KS equations are:
\begin{equation}
    \bigg\{ - \frac{\hbar^2}{2m}\nabla^2 + v_{\text{KS}}[\rho](x,\sigma) \bigg\}\varphi_j(x,\sigma) = \epsilon_j \varphi_j(x,\sigma),
    \label{eqn:KSeqnPIB}
\end{equation}
where the set of spin-polarized wavefunctions $\{\varphi_j : j \geq 1\}$ are the solutions ordered by energy. The KS many-body wavefunction $\Psi$, generally assumed to be a single Slater determinant, is built from $\{\varphi_j\}$. Both $\{\varphi_j\}$ and their corresponding energies $\{\epsilon_j : j \geq 1\}$ are typically obtained iteratively from self-consistent field (SCF) calculations, but in this case, because we have set $v_{\text{KS}}(x,\sigma) = 0$, we know $\epsilon_j$ exactly from the analytical solutions to the non-interacting PIB problem and do not need to find the solutions through minimization:
\begin{equation}
    \epsilon_n = \frac{n^2 \pi^2 \hbar^2}{8 m_e L^2},
    \label{eqn:pibe}
\end{equation}
where $m_e$ is the mass of an electron. For the same reason, we know the wavefunctions solving equation (\ref{eqn:KSeqnPIB}) exactly, without the need for minimization, which are defined by their quantum number $n$:
\begin{equation}
    \phi_n(x) = \sqrt{\frac{1}{L}}\text{sin}\bigg( \frac{n \pi}{2L} x\bigg), \hspace{2em} (n = 2, 4, ...),    \label{eqn:pibphieven} 
\end{equation}
\begin{equation}
    \phi_n(x) = \sqrt{\frac{1}{L}}\text{cos}\bigg( \frac{n \pi}{2L} x\bigg), \hspace{2em} (n = 1, 3, ...).    \label{eqn:pibphiodd}
\end{equation}

From each spatial wavefunction, $\phi(x)$, one can form two different orthonormal spin and space-dependent wavefunctions by multiplying the spatial function by the up $\alpha(\sigma)$ or down $\beta(\sigma)$ spin function: \cite{alma9916908026606531}
\begin{equation}
\varphi(x,\sigma) =
    \begin{cases}
            \phi(x)\alpha(\sigma)   \\
              \hspace{1.5em}  \text{or} \\
            \phi(x)\beta(\sigma),
    \end{cases}
    \label{eqn:spinfunctions}
\end{equation}
The density for a system of non-interacting particles is:
\begin{equation}
    \rho(x,\sigma) = \sum^{\infty}_{j=1} f_j |\varphi_j(x,\sigma)|^2,
    \label{eqn:rho}
\end{equation}
with occupations $f_j \in \{0, 1\}$ to specify occupied and unoccupied states.
Up to two $\varphi_j$ may correspond to the same $\phi_n$, which is the case for a doubly occupied spatial state. 
Knowing the non-interacting density, the sum of the wavefunction energies, the Hartree energy and an approximation to the XC energy functional, the total interacting energy is obtained as\cite{Kaxiras}
\begin{equation}
    E^{\text{tot}}[\rho] = \sum^{\infty}_j f_j \epsilon_j - E_H[\rho] - \\ \sum_{\sigma=\alpha}^\beta \int \big(v_{\text{xc}}[\rho](x,\sigma)\big)\rho(x,\sigma) dx  + E_{\text{xc}}[\rho],
    \label{eqn:EtotGS}
\end{equation}
where $\frac{\delta E_{\text{xc}}[\rho]}{\delta \rho(x,\sigma)} = v_{\text{xc}}[\rho](x,\sigma)$. Equation (\ref{eqn:EtotGS}) is exact if the XC functional is known exactly.

\subsection{\label{subsec:EDFT}Ensemble Density Functional Theory}
EDFT as discussed here stems from Theophilou and Gidopoulos's work in 1987 which built ensembles from KS states.\cite{T87} This variational principle for equi-ensembles was generalized to ensembles of monotonically decreasing, non-equal weights by Gross-Oliveira-Kohn (GOK) in 1988. \cite{GOK88I}
To avoid confusion, we note that the theory of thermal \enquote{Mermin} DFT,\cite{PhysRev.137.A1441} commonly used for periodic systems such as metals, has been referred to as \enquote{ensemble DFT} also,\cite{PhysRevLett.79.1337} but it is based on a Fermi-Dirac thermal ensemble and thus is quite different from GOK EDFT.

\begin{figure}
    \centering
    \includegraphics[width=120mm,scale=1]{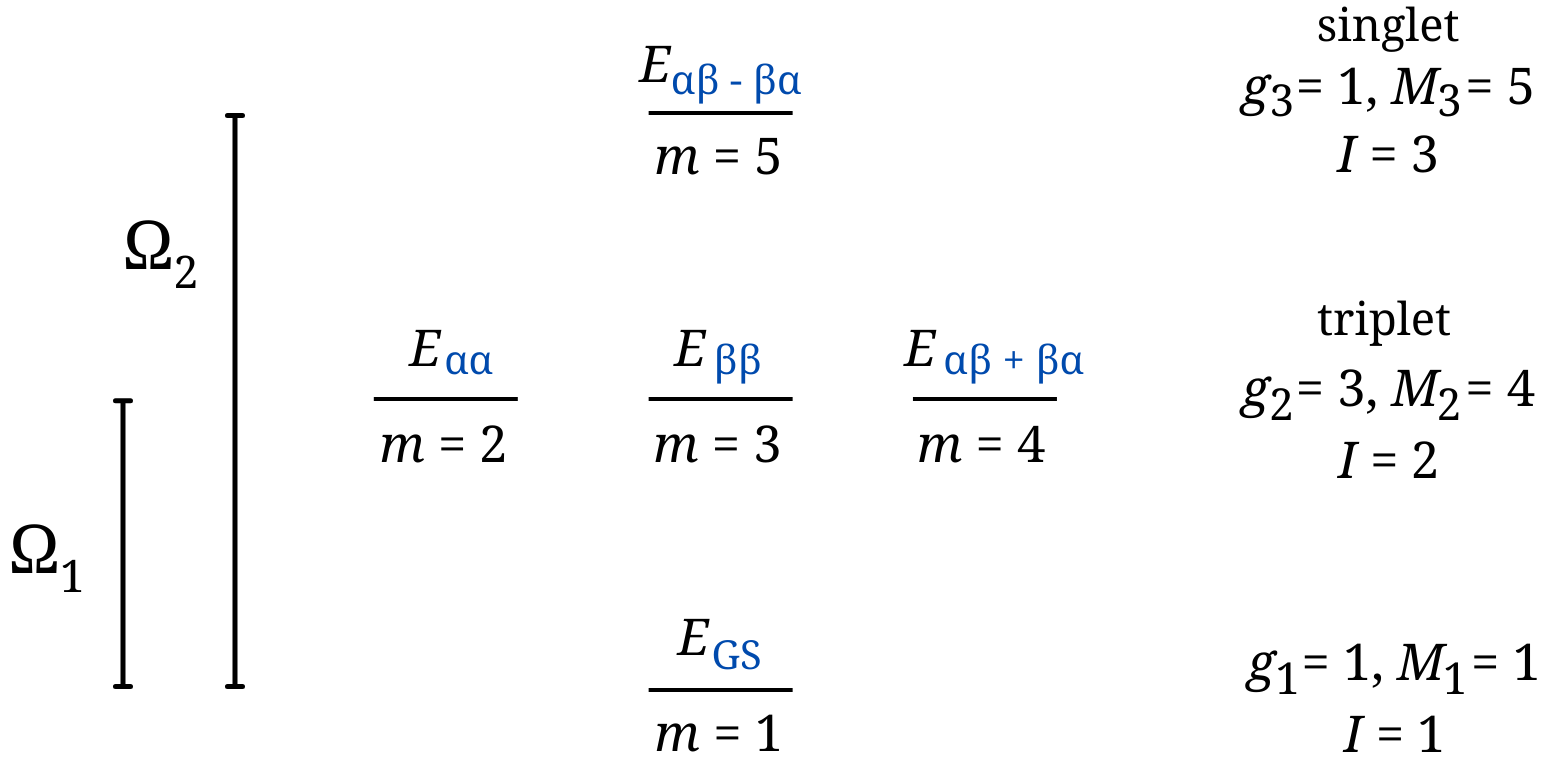}
    \caption{Diagram of the multiplet structure for the ensemble of interacting particles for a PIB, obeying spin symmetry. $I=2$ corresponds to the bi-ensemble, which includes up to $m=4$ (2nd multiplet). $I=3$ corresponds to the tri-ensemble, which includes up to $m=5$ (3rd multiplet). The degeneracy of the highest multiplet included in the ensemble is given by the corresponding value of $g_I$, and $M_{I}$ is the total number of states included in the $I$th ensemble. The assignment of $m$'s within a multiplet are arbitrary.}
    \label{fig:mult}
\end{figure}

The ensemble-generalized form of equation (\ref{eqn:KSeqnPIB}) is the non-interacting ensemble KS equation:
\begin{equation}
\left\{-\frac{1}{2}\nabla^2 + v^w_{\text{KS}}[\rho^w](x,\sigma)\right\}\varphi^w_{j}(x,\sigma) = \epsilon^w_{j}\varphi^w_{j}(x,\sigma),
\label{eqn:ensKS}
\end{equation}
where $\varphi^w_{j}$ are the non-interacting single-particle wavefunctions that reproduce the ensemble density, $\rho^w(x,\sigma)$. The KS many-body wavefunctions $\{\Psi_m^{w}[\rho] : 1 \leq m \leq M_{I}\}$, assumed to be Slater determinants or linear combinations of Slater determinants, are built from $\{\varphi^w_j(x,\sigma) : j \geq 1\}$ having individual energies $\epsilon^w_j$ which are obtained from the ensemble KS equation, equation (\ref{eqn:ensKS}). Symmetry-adapted linear combinations of Slater determinants may be used, as the conventional restriction to single Slater determinants has been found to be overly restrictive in EDFT.\cite{GP17} The ensemble-generalized form of equation (\ref{eqn:vs}) is:
\begin{equation}
v^w_{\text{KS}}[\rho^w](x,\sigma) = v_{\text{ext}}(x) + \frac{\delta E^w_{\rm Hxc}[\rho^w]}{\delta \rho^w(x,\sigma)},
\end{equation}
and the ensemble functional for Hartree, exchange, and correlation (HXC), $E^w_{\rm Hxc}$, may be separated into its constituent parts:

\begin{equation}
    \frac{\delta E^w_{\rm Hxc}[\rho^w]}{\delta \rho^w(x,\sigma)} = \int \frac{ \sum_{\sigma=\alpha}^\beta \rho^w(x', \sigma)}{|x-x'|}dx' + \frac{\delta E_{\rm x}^w[\rho^w]}{\delta \rho^w(x,\sigma)} + \frac{\delta E_{\rm c}^w[\rho^w]}{\delta \rho^w(x,\sigma)}.
\end{equation}
While the single-particle wavefunctions $\{\varphi^w_{j}\}$ and their corresponding energies $\{\epsilon_j\}$ are calculated in the same way as in the GS case presented in section \ref{sec:1Dwell}, the ensemble density is constructed in a different way than the GS equation (\ref{eqn:rho}):

\begin{equation}
\rho^w(x,\sigma) = \sum^{M_{I}}_{m=1} \mathtt{w}_m \bigg(\sum^\infty_{j=1} f^{m}_{j}|\varphi^w_j(x,\sigma)|^2 \bigg),
\label{eqn:ensdens}
\end{equation}
where $f^m_{j}$ denotes the occupation of $\varphi^w_j(x,\sigma)$ in the $m$th KS wave function $\Psi^w_m[\rho^w]$.\cite{MSFL20} $I$ denotes the set of degenerate states (or \enquote{multiplet}) with the highest energy in the ensemble. This set can be equivalently referred to as the $(M_{I})$th state, as we consider an \hide{$I$th-order} ensemble of $M_{I}$ (possibly degenerate) electronic states each consisting of $N_e$ electrons, numbered from $m=1$ to $M_{I}$. Then, $g_I$ is the multiplicity of the $I$th multiplet, and $M_{I}$ is the total number of states up to and including the $I$th multiplet, $M_I = \sum_{j=0}^{I}g_j$.\cite{GOK88I, MSFL20} $I=2$ denotes a bi-ensemble, and $I=3$ denotes a tri-ensemble, as depicted in figure \ref{fig:mult}.

GOK ensembles must include all of each degenerate subspace to be well-defined. Each many-electron state's energy is denoted by $E_{m=1} \leq ... \leq E_{m=M_{I}}$, and the energy of the $m$th KS state is
\begin{equation}
E_m = \sum^\infty_{j=1} f^{m}_{j} \epsilon_j,
\label{eqn:EmthKS}
\end{equation}
which can be obtained exactly in this case from equation (\ref{eqn:pibe}). %
Each state is assigned a weight $\mathtt{w}_m$ from the set $\{w\} \equiv (\mathtt{w}_{m=1}, ..., \mathtt{w}_{m=M_{I}})$ of monotonically non-increasing $(\mathtt{w}_{m=1} \geq ... \geq \mathtt{w}_{m=M_{I}})$ weights obeying
\begin{equation}
    \sum_{m=1}^{M_{I}} \mathtt{w}_m = 1.
\end{equation}
For the GOK ensembles considered here, the weights are defined as\cite{GOK88I}
\begin{equation}
\mathtt{w}_m =
    \begin{cases}
            \frac{1-\mathtt{w}g_I}{M_{I} - g_I} & m \leq M_{I} - g_I,   \\
            \mathtt{w} &   m > M_{I} - g_I, 
    \end{cases}
    \label{eqn:GOKweights}
\end{equation}
where $\mathtt{w} \in [0, 1/M_{I}]$, such that all states but those in the highest ($I$th) multiplet have the same weight, and only the GOK weight, $\texttt{w}$, is needed to define the weighting of the ensemble. By definition, $\mathtt{w} = \mathtt{w}_{M_{I}}$. 
The total ensemble energy \cite{GOK88I} is approximated as
\begin{equation}
\mathcal{E}^w[\rho^w] = \sum_{m=1}^{M_I} E_m  - E_{\text{H}}[\rho^{w}] \\ - \sum_{\sigma=\alpha}^\beta \int \bigl(v_{\text{xc}}[\rho^w](x,\sigma)\bigr)\rho^w({x,\sigma})d^3x  \\ + E_{\text{x}}^{\text{LDA}}[\rho^w] +
E_{\text{c}}^{\text{LDA}}[\rho^w].
\label{eqn:Etotens}
\end{equation}
The exact ensemble energy would be obtained if the ensemble HXC functional were used. \hide{In this setup \todo{which setup, GOK?}, Degenerate states are always assigned equal weights. }

In the case of our bi-ensemble in which KS eigenvalues are spin-independent, we differentiate equation (\ref{eqn:Etotens}) with respect to $\mathtt{w}$, as in reference \citenum{GOK88I} but considering states which are not necessarily single Slater determinants and allowing $f_j^m$ to be 0, 1/2, or 1. Then we obtain the first excitation energy from the GS: %
\begin{equation}
    \Omega_{1} = 
    \epsilon_{n=N_e/2+1}-\epsilon_{n=N_e/2}
    + \left. \frac{1}{3} \frac{\partial E_{\text{Hxc}}^{w, I=2} [\rho]}{\partial \texttt{w}} \right|_{\rho=\rho^w_{I=2}} ,
    \label{eqn:omega_1}
\end{equation}
The third term on the right of equation (\ref{eqn:omega_1}) is the \enquote{ensemble correction} to the non-interacting difference of energies %
from equation (\ref{eqn:EmthKS}). %
For the tri-ensemble, one obtains from equations (100) and (89) of reference \citenum{GOK88I}: %
\begin{align}
    \Omega_{2} &= \frac{1}{g_3}\left.\frac{d\mathcal{E}^{w}_{I=3}}{d\mathtt{w}}\right|_{\mathtt{w}=\mathtt{w}_{I=3}} +\frac{1}{M_2}\left.\frac{d\mathcal{E}^w_{I=2}}{d\mathtt{w}}\right|_{\mathtt{w}=\mathtt{w}_{I=2}}\notag\\
    &=\frac{1}{g_3} \left(\sum_{j=1}^\infty\left(\sum_{m=M_2 + 1}^{M_3} f_j^m \epsilon_j-\frac{g_3}{M_2}\sum_{m=1}^{M_2} f_j^m\epsilon_j \right) +\left.\frac{\partial E_{Hxc}^{w,I=3}[\rho]}{\partial \mathtt{w}}\right|_{\rho=\rho^w_{I=3}} \right)\notag\\
    &+\frac{1}{M_2} \left(\sum_{j=1}^\infty\left(\sum_{m=M_1 + 1}^{M_2} f_j^m \epsilon_j-\frac{g_2}{M_1}\sum_{m=1}^{M_1} f_j^m\epsilon_j \right) +\left.\frac{\partial E_{Hxc}^{w,I=2}[\rho]}{\partial \mathtt{w}}\right|_{\rho=\rho^w_{I=2}} \right).
    \label{eqn:omega_2}
\end{align}
For our specific tri-ensemble, as in figure \ref{fig:mult}, the excitation energy for $I=3$ can be written in terms of the individual KS orbital energies as:
\begin{align}
    \Omega_2^{\rm}&=\frac{1}{4}\left(\epsilon_{n=N_e/2+1}^{w_2}-\epsilon_{n=N_e/2}^{w_2} \right)\notag
    +\frac{3}{4}\left(\epsilon_{n=N_e/2+1}^{w_1}-\epsilon_{n=N_e/2}^{w_1} \right)\notag\\
    &+
    \left.\frac{\partial E_{Hxc}^{w,I=3}[\rho]}{\partial \mathtt{w}}\right|_{\rho=\rho^w_{I=3}}+\frac{1}{4}\left.\frac{\partial E_{Hxc}^{w,I=2}[\rho]}{\partial \mathtt{w}}\right|_{\rho=\rho^w_{I=2}}, 
\end{align}
where superscripted $w_i$ indicate that these are the weight-dependent KS eigenvalues for the ensemble including up to the $I=i$ multiplet. (Note that it is essential for such expression to be not just a linear combination of eigenvalues but a linear combination of eigenvalue \textit{differences}, to avoid any dependence on an overall energy offset.) We now assume that the eigenvalues of our KS system for the bi-ensemble and the tri-ensemble have negligible differences, consistent with our assumption that the PIB states are our ensemble-KS states for the minimizing ensemble density, $\rho^w$. Then we can further simplify to the approximate expression in terms of the eigenvalues of the tri-ensemble KS system,
\begin{align}
    \Omega^{\rm}_2&\approx \epsilon_{n=N_e/2+1}^{w_2}-\epsilon_{n=N_e/2}^{w_2}+
    \left.\frac{\partial E_{\text{Hxc}}^{w,I=3}[\rho]}{\partial \mathtt{w}}\right|_{\rho=\rho^w_{I=3}}+\frac{1}{4}\left.\frac{\partial E_{\text{Hxc}}^{w,I=2}[\rho]}{\partial \mathtt{w}}\right|_{\rho=\rho^w_{I=2}}. 
    \label{eqn:final_omega_2}
\end{align}

\subsection{\label{subsec:HXC}Approximations to Hartree, Exchange, and Correlation}

\hide{\todo{Look at these refs}\cite{FHMP98, LN99, T79, T87, GP17}}

The development of accurate weight-dependent density-functional approximations (DFAs) for EDFT is an ongoing challenge. 
Existing ensemble approximations to $E_{\text{xc}}^w$ include the quasi-local-density approximation (qLDA) functional,\cite{GOK88II, PhysRevA.34.737} the \enquote{ghost}-corrected exact exchange (EXX) functional,\cite{N98, GPG02} the exact ensemble exchange functional (EEXX),%
\cite{https://doi.org/10.1002/qua.560560833} local system-dependent and excitation-specific ensemble exchange functionals for double excitations (CC-S)\cite{MSFL20} and for single excitations,\cite{LF20} a universal weight-dependent local correlation functional (eVWN5) based on finite UEGs, \cite{MSFL20} and the orbital-dependent second-order perturbative approximation (PT2) for the ensemble correlation energy functional.\cite{PhysRevA.104.052806, CPG22}
As noted in all the aforementioned works, ensemble HXC has  special complications beyond those of GS DFT, such as the consideration that ensemble Hartree and exchange are not naturally separated in EDFT.\cite{GSP20}
Though each of these approaches above to approximating ensemble XC energies provides insight into the necessary characteristics of ensemble DFAs, it is unclear whether any of them are appropriate for periodic systems since they were developed for localized systems. In this work, for a first exploration of EDFT on periodic systems, we choose a simple approximation based on a Local Spin Density Approximation (LSDA).

The \enquote{traditional} DFAs of GS DFT can be used for ensembles by evaluating them on ensemble densities:
\begin{equation}
E_{\text{Hxc}}^\text{trad}[\rho] = E_{\text{Hxc}}\bigg[\sum_{m=1}^{M_{I}} \mathtt{w}_m \rho_\textit{m}\bigg].
\label{eqn:Ans1}
\end{equation}
This use of the ensemble density with GS DFAs, typically only applied to Hartree and exchange, has been called \enquote{\textit{Ansatz} 1.}\cite{GSP20} The use of ensemble densities in \enquote{traditional} GS DFAs results in fictitious interactions of ground- and excited-state densities, or \enquote{ghost interaction errors} (GIEs), in both Hartree and exchange which do not cancel each other. \cite{GSP20, GPG02, PP14, GOK88II, PYTB14} %
A DFA for open-shell systems which is free of ghost-exchange-error approach has also been proposed to address this issue. \cite{CPG22}Additionally, with this form of ensemble DFA, the derivatives in equation (\ref{eqn:omega_1}) become zero, since the weight dependence is within the ensemble density only. As such, nothing is learned from application of EDFT in such an approximation. We instead opt to use ensemble-generalized LSDA, in which we build an ensemble average by evaluating the GS Hartree and LSDA functionals on the density of each state in the ensemble individually:
\begin{equation}
E_{\text{Hxc}}^{\text{LSDA},w}[\rho] = \sum_{m=1}^{M_{I}} \mathtt{w}_m E_{\text{Hxc}}^{\text{LSDA}}[\rho_\textit{m}],
    \label{eqn:Hxc}
\end{equation}
which has been called \enquote{\textit{Ansatz} 2.}\cite{GSP20} In this way, we ensure the ensemble functionals are weight-dependent, giving us nonzero corrections in equation (\ref{eqn:omega_1}). While in this work we do not need to evaluate the ensemble energy variationally since the exact KS solutions are known by construction, we note that the variational evaluation of the ensemble energy with equation \ref{eqn:Hxc} is complicated because it does not lead to conventional KS equations.\cite{GK21}

Derivatives of this equation with respect to $\mathtt{w}$ depend on the weights defined in equation (\ref{eqn:GOKweights}), which in turn are determined by the multiplet structure and $I$, e.g., whether a bi-ensemble or tri-ensemble is used  (figure \ref{fig:mult}), and have the general form:

\begin{equation}
     \left. \frac{\partial E^{\text{LSDA},w,I}_{\text{Hxc}} [\rho]}{\partial \mathtt{w}} \right|_{\rho=\rho^w}
      = - \frac{g_I}{M_{I} - g_I} \sum_{m=1}^{M_{I}-g_I} E_{\text{Hxc}}[\rho_\textit{m}] \\
      + \sum_{m=M_{I}-g_I+1}^{M_{I}} E_{\text{Hxc}}[\rho_\textit{m}] .
      \label{eqn:general_correction}
\end{equation}
Note a useful property: the sum of the coefficients of the $M_I$ states is
\begin{align}
- \frac{g_I}{M_{I} - g_I} \left( M_{I} - g_I \right) + \left( 1 \right) g_I = -g_I + g_I = 0 .
\label{eqn:sum_to_zero}
\end{align}
This property is essential allow the excitation energy be intensive (size-consistent) as we approach the thermodynamic limit, since individual total energy terms are extensive and grow without bound.

While the true interacting wavefunctions have no weight-dependence, the KS wavefunctions, and consequently the KS state densities, are weight-dependent. This weight-dependence results in principle in an additional term in the HXC derivatives in our equations for $\Omega$:
\begin{equation}
    \left. \frac{\partial E_{\text{Hxc}}^w [\rho]}{\partial \texttt{w}} \right|_{\rho=\rho^w} = \sum_{m=1}^{M_I} \left\{\frac{\partial \mathtt{w}_m}{\partial \mathtt{w}} E_{\text{Hxc}}\left[\rho_m\right] + w_m \sum_{\sigma=\alpha}^\beta \int 
    \frac{\partial \rho_m^{w}(x,\sigma)}{\partial \mathtt{w}} dx \left. \frac{\partial E_{\text{Hxc}}\left[\rho\right]}{\partial \rho(x,\sigma)} \right|_{\rho=\rho_m} \right\}
\end{equation}
Practically, we neglect this weight-dependence which is related to taking derivatives at constant $\rho$. This is consistent with our construction of a system with $v_{\text{KS}}\left( x, \sigma \right) = 0$ for the ground state, for which no SCF calculations are needed. We assume that $v_{\text{KS}}\left( x, \sigma \right) = 0$ for all excited states as well, which cannot necessarily be satisfied by construction. Note, however, that in the thermodynamic limit of the 1D UEG, the densities of all states are identical (and uniform), and the spin densities of singlet excited states are also identical to the ground state, implying identical $v_{\text{KS}}$ for the singlet states. Therefore the approximation should improve in the thermodynamic limit.  %

The definition of ensemble-generalized Hartree in equation (\ref{eqn:Hxc}) is GIE-free.\cite{PYTB14} Though this choice avoids a significant source of GIE, our current form of ensemble-generalized LSDA does introduce some GIE from XC.\cite{GSP20} 
We report results for ensemble corrections which have been built using the weight-dependent Hartree of equation (\ref{eqn:Hxc}), denoted by HXC, and also for the case where there is no Hartree contribution to the correction, denoted by XC, due to the \enquote{traditional} Hartree definition in equation (\ref{eqn:Ans1}). 

\subsection{\label{subsec:Slater}Densities of Ground and Excited States}

Here we show explicitly the spin-polarized densities involved in the ground and excited states which we use in our EDFT calculations. %
All densities involved here include a contribution from the closed shell,
\begin{equation}
    \rho_{\rm closed}(x,\sigma) = \sum^{N_e/2-1}_{n=1} |\phi_n(x,\sigma)|^2 \left(|\alpha(\sigma)|^2 + |\beta(\sigma)|^2 \right) ,
\end{equation}
and the ground-state density is
\begin{equation}
\rho_{\rm GS}(x,\sigma) = \left| \phi_1(x) \right| \left( \left|\alpha(\sigma)\right|^2 +  \left|\beta(\sigma)\right|^2 \right) + \rho_{\rm closed}(x,\sigma) ,
\end{equation}
where $\phi_1$ is the highest occupied state.

In the spin-polarized PIB system of even $N_e$, based on spin symmetry, the system has a nondegenerate GS, a triplet first excited state, and a singlet second excited state, as depicted in figure \ref{fig:mult}. An odd number of $N_e$ would result in a different multiplet structure, but we do not investigate that case here, since odd/even distinctions should disappear in the thermodynamic limit anyway. The density of the $\alpha \alpha$ state ($m_s = 1$) in the triplet, obtained from its Slater determinant and then written in terms of its constituent wavefunctions, is:
\begin{equation}
\rho_{\alpha \alpha}(x,\sigma) = \left| \phi_1(x) \alpha(\sigma)\right|^2 + \left| \phi_2(x) \alpha(\sigma)\right|^2 + \rho_{\rm closed}(x,\sigma),
\label{eqn:upup}
\end{equation}
where $\phi_2$ is the lowest unoccupied state, with reference to the ground state.
Then, for the $\beta\beta$ ($m_s = -1$) state in the triplet, we obtain a similar equation where the $\alpha$ spins are flipped to $\beta$ spins:
\begin{equation}
\rho_{\beta \beta}(x,\sigma) = \left| \phi_1(x) \beta(\sigma)\right|^2 + \left| \phi_2(x) \beta(\sigma)\right|^2 + \rho_{\rm closed}(x,\sigma).
\label{eqn:dndn}
\end{equation}
While $\rho_{\alpha \alpha}(x,\sigma) \neq \rho_{\beta \beta}(x,\sigma)$, our approximations to the energy-density functional, evaluated on these two densities, yields the same numerical result for their energies as required by symmetry, and is the result obtained from any LSDA. For the $m_s = 0$ excited states, we must use linear combinations of two Slater determinants to obtain the density: 
\begin{equation}
     \rho_{\alpha \beta \pm \beta \alpha}({x,\sigma}) = \frac{1}{2} \bigg( |\phi_1(x)\alpha(\sigma)|^2 + |\phi_1(x)\beta(\sigma)|^2 + \\ |\phi_2(x)\alpha(\sigma)|^2 + |\phi_2(x)\beta(\sigma)|^2 \bigg) + \rho_{\rm closed}(x,\sigma).
     \label{eqn:Ms0dens}
\end{equation}
We obtain the same density for the symmetric triplet ($+$) and antisymmetric singlet ($-$) sum of the two Slater determinants. No pure density functional can tell the two states, having identical densities, apart, despite the fact that the triplet should be degenerate with the other two triplet states.

We will consider later, in sections \ref{sec:Bi_e} and \ref{sec:Bi_b}, two approaches to treating the triplet energy. The first method, outlined in \ref{sec:Bi_b}, is the symmetry-broken bi-ensemble, in which the correct densities for each state in the triplet, obtained from equations  (\ref{eqn:upup}), (\ref{eqn:dndn}), and (\ref{eqn:Ms0dens}), are used. Although $E_{\rm xc}[\rho_{\alpha \alpha}]$ and $E_{\rm xc}[\rho_{\beta \beta}]$ are equal, the energy $E_{\rm xc}[\rho_{\alpha \beta + \beta \alpha}]$ of the third member of the triplet has a higher energy in LSDA, with their difference decreasing asymptotically towards 0 as $N_e \rightarrow \infty$. %
Since both the singlet and triplet $m_s = 0$ states have the same density, we will write $\rho_{\alpha \beta \pm \beta \alpha}$ to refer to their shared density. To address this symmetry-breaking issue, we have also considered the symmetry-enforced bi-ensemble in section \ref{sec:Bi_e}, in which we do not use the computed value of $E[\rho_{\alpha \beta + \beta \alpha}]$ at all, and instead use the value of $E[\rho_{\alpha \alpha}] = E[\rho_{\beta \beta}]$ to represent all three states, maintaining the degeneracy of the KS states forming the spin triplet. %

GOK ensemble theory requires that states are ordered based on the energies of the interacting system, and that all states from the GS up to and including the $I$th multiplet are included in the ensemble.
It is not always practically feasible to be certain that there are no additional states lying between those we have included in the system,\cite{MSFL20} but we work under the assumption that we have included all states between the ground state and $I$th excited state, such that we have not violated the rules of the GOK ensemble. 
\begin{figure}
    \centering
    \includegraphics[width=120mm,scale=0.48]{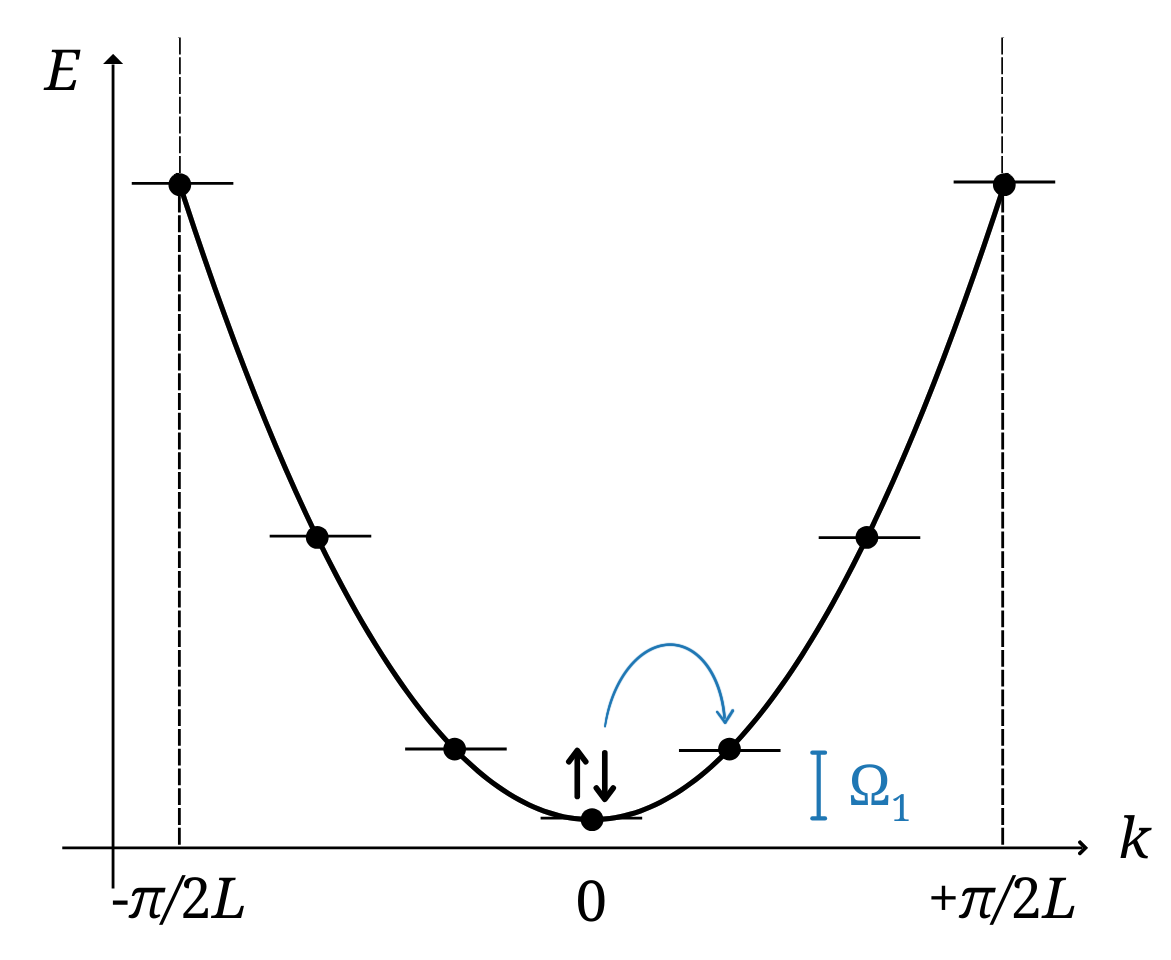}
    \caption{Band structure of the free electron gas, shown with a discrete set of
    6 $k$-points in a Brillouin zone based on a periodic length $2L$. Arrow depicts an excitation with $\Delta k > 0$.}
    \label{fig:kpoints}
\end{figure}
\subsection{\label{sec:UEG}1D Uniform Electron Gas}
We first consider a possible way that a periodic system could be discretized to allow application of EDFT. Leaving aside the question of whether such an approach is theoretically sound, we find that in the case of the UEG (the limit of our model) corrections to the KS excitation energies are identically zero, demonstrating that alternate strategies are needed in order to obtain non-trivial results. 

We consider an infinite limit of our system in which the KS potential is zero everywhere, and periodic boundary conditions $\phi(x+2L) = \phi(x)$ are imposed for an arbitrary repeating cell of length $2L$. The KS wavefunctions have the form 
\begin{equation}
    \phi_k(x) = \sqrt{\frac{1}{2L}}e^{ik x} ,
    \label{eqn:singleelectronkpointwfn}
\end{equation}
and the KS energies for such a 1D system are
\begin{equation}
E = \frac{\hbar^2 k^2}{2m_e},
\label{eqn:kEnergy}
\end{equation}
as discussed further in section \ref{sec:effmass}. 

This system has a continuous spectrum of states and, as noted earlier, the GOK EDFT has been defined only for a discrete spectrum. We consider, consistent with our boundary conditions, a set of $k$-points $k = (\pi n/ L)$ where $n = \left\{ 0, \pm1, \pm2, 3 \right\}$. This discretization is equivalent to construction of a \enquote{finite, but topologically periodic system,} like a particle on a ring,\cite{KK14} such as one might construct to avoid the edge effects of our finite PIBs.
We consider an excitation with $\Delta k > 0$, to keep things simple and involve only one excited KS energy level -- while this bends the rules of the GOK EDFT by not assigning the same weights to all of a degenerate set, it can be justified in a generalization in which states of different symmetry (e.g. crystal momentum $k$) can be treated separately.\cite{AJC}  With the two KS energy levels, we obtain a singlet-triplet structure which is the same as in our finite well with even $N_e$ (Section \ref{sec:1Dwell} and figure \ref{fig:mult}). %
Filling the system with 2 electrons per cell of $2L$ results in two electrons in the lowest $k$-point, as in figure \ref{fig:kpoints}. %
With two electrons per unit cell, moving an electron from one $k$-point to the next represents exciting $1/2$ of all electrons in the periodic system. All of the ground- and excited-state densities are constant; e.g. from equation (\ref{eqn:upup}) we obtain:

\begin{equation}
    \rho_{\alpha \alpha}(x,\sigma) = \bigg|\sqrt{\frac{1}{2L}}e^{ik_1 x}\alpha(\sigma)\bigg|^2 + \bigg|\sqrt{\frac{1}{2L}}e^{ik_2 x}\alpha(\sigma)\bigg|^2 \\ = \frac{1}{L},
    \label{eqn:kGS}
\end{equation}
where $k_2 = k_1 + \Delta k$. We find the same result for the three states which make up the triplet of the first excited state, equations (\ref{eqn:Ms0dens}), (\ref{eqn:upup}), and (\ref{eqn:dndn}), and for the GS.
The energy correction, as will be shown in Section \ref{sec:Bi_e}, is
\begin{equation}
\left. \frac{\partial E_{\text{Hxc}}^{w, I=2} [\rho]}{\partial \mathtt{w}} \right|_{\rho=\rho^w_{I=2}} = -3 E_{\text{Hxc}}[\rho_{\text{GS}}] + E_{\text{Hxc}}[\rho_{\alpha \alpha}] \\ + E_{\text{Hxc}}[\rho_{\beta \beta}] + E_{\text{Hxc}}[\rho_{\alpha \beta \pm \beta \alpha}] = 0,
\label{eqn:correctionline}
\end{equation}
where each density is identical,
and the total correction goes to zero because the coefficients in front of each energy term always sum to zero (equation (\ref{eqn:sum_to_zero})). %
Since the GOK ensemble correction depends on the ensemble density defined in equation (\ref{eqn:ensdens}), and each state has the same density, it is not possible to obtain a non-zero correction from EDFT to the UEG in this manner. 
Changing the number of electrons, number of $k$-points, length of the box, or which excitation we calculate (e.g. including $\Delta k < 0$) would change the complexity for this model, but not the basic conclusion. We instead study a finite system which increases in size towards the thermodynamic limit to gain information about the behavior of EDFT's correction as it approaches a periodic system.
\subsection{\label{subsec:Limit}Thermodynamic Limit of the Finite-Length Well}
\begin{figure}
\includegraphics[width=120mm,scale=0.50]{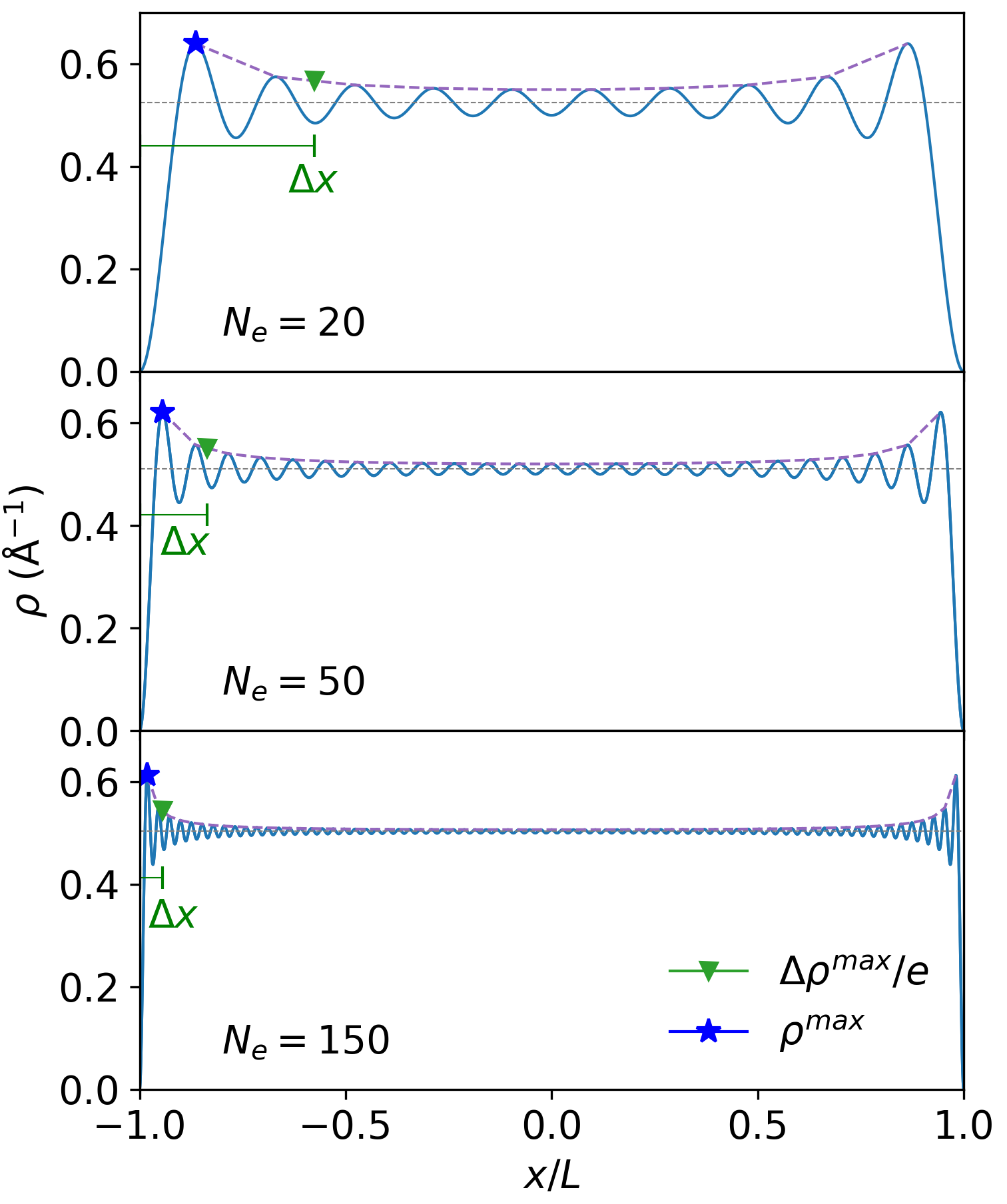}
\caption{\label{fig:deltax} The non-interacting ensemble densities for $N_e = 20$, $N_e = 50$, and $N_e = 150$, with interpolation line (dashed) between peaks. The maximum of the density within $-L \leq x \leq 0$ is $\rho^{\text{max}}$. The average density
at the center of the box is shown as a dotted horizontal line. The width of the edge region, $\Delta x$, is defined as the width from the edge of the box to the point at which the interpolation line is $\Delta \rho^{\max}/e$ from the center peak-to-peak amplitude average, as described in section \ref{subsec:Limit}. The width of $\Delta x$ spans a smaller portion of the box as $N_e$ increases.}
\end{figure}

\begin{figure}
\includegraphics[width=120mm,scale=0.5]{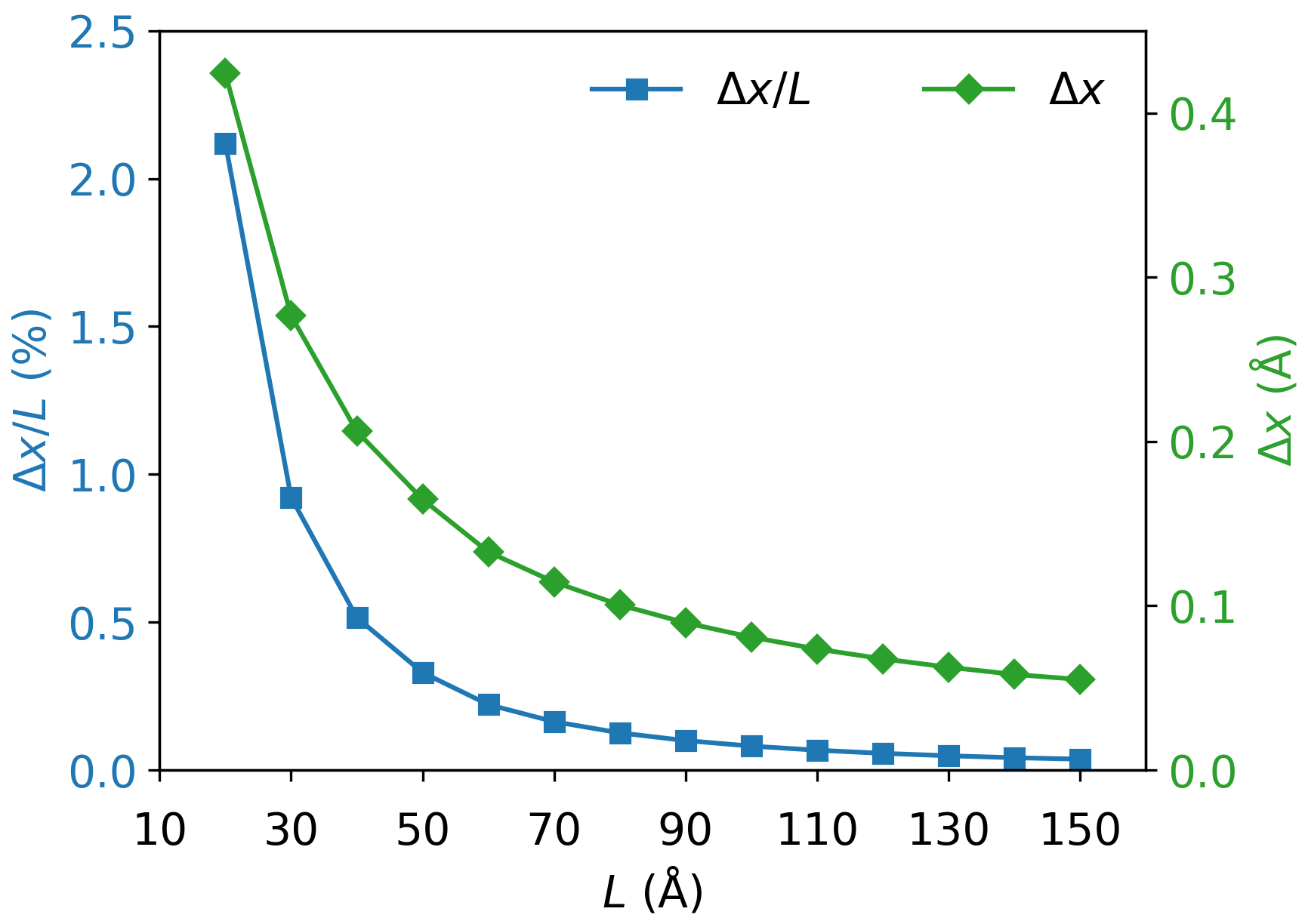}%
\vspace{-0.40cm}\caption{\label{fig:nonint300} Decreasing width of the edge regions of the density, with increasing $L$. The average density is held constant to 0.5 $\text{\AA}^{- 1}$ as in equation (\ref{eqn:avgdens}). The right axis is the width of the region $\Delta x$, and the left axis is the percentage of the half-length of the box spanned by $\Delta x$.}
\end{figure}

We increase the number of electrons in our system along with the length of the box, holding the average density constant:
\begin{equation}
\frac{N_e}{2L} = 0.5\ \text{\AA}^{-1} \hspace{3em} N_e,L \rightarrow \infty.
\label{eqn:avgdens}
\end{equation}
As $N_e \rightarrow \infty$, a region of increasingly constant density begins to form at the center of the box, with decreasing oscillations and decreasing edge regions. According to the Wentzel–Kramers–Brillouin (WKB) Approximation, there will always be a peak at the classical turning points,\cite{Sakurai} i.e. the edges of the box. As both $N_e$ and $L$ approach infinity, the density of the system becomes more uniform, with the nonuniform edge regions decreasing in width. To quantify this property (figure \ref{fig:deltax}), we first find the height $\rho^{\text{max}}$ of the highest peak within $-L \leq x \leq 0$. We average the values of the peaks and troughs of the density at the center ($x = 0$) to find the average uniform density $\rho^{\text{uniform}}$. We then define $\Delta \rho^{\text{max}} = \rho^{\text{max}} - \rho^{\text{uniform}}$. Next, we consider an envelope function that excludes the oscillations of the density by linearly connecting the peaks of the density. We determine the width $\Delta x$ of the region between the edge of the box and the position at which the envelope has decreased to $\Delta \rho^{\text{max}}/e$ measured from $\rho^{\text{uniform}}$. We note that $\Delta x$ decreases not only as a fraction of $L$ but also in absolute terms, demonstrating that our model becomes increasingly uniform  with increasing $L$ and that edge effects become negligible (figure \ref{fig:nonint300}). 
In this way, our model systems in the approach to the thermodynamic limit can be used to study how EDFT %
performs in a uniform periodic system. 
\section{\label{sec:Methods}Computational Methodology}
Octopus is uniquely suited for this work due to its ability to define arbitrary potentials and therefore easily treat model systems and 1D systems.\cite{Octopus_PCCP,Tea20} In this work we use Octopus version 11.4. In order to realize our condition of setting the KS potential equal to the 1D finite well potential in Octopus, the potential is set to zero within a finite domain determined by $L$. The wavefunction is constrained to zero at the boundaries of the box. %
We limit our system to an even number of electrons $N_e$ whose ratio to $L$ is held fixed as in equation (\ref{eqn:avgdens}), and consider its spin-polarized solutions obtained from the PIB as in equations (\ref{eqn:pibphieven}), (\ref{eqn:pibphiodd}) and (\ref{eqn:rho}).  The starting initial guesses in the Kohn-Sham equations are random wavefunctions. We used the conjugate-gradients eigensolver with a tolerance of $10^{-6}$ eV, which can require up to 1000 eigensolver iterations, and did not use a preconditioner. Eigensolver convergence was difficult to achieve and we settled on this fixed density ratio, grid, and the eigensolver to give adequate convergence behavior. The average density of 0.5 $\text{\AA}^{\minus 1}$ was used to achieve eigensolver convergence since systems with the larger average density of 1 $\text{\AA}^{\minus 1}$ were unable to be converged for all values of $N_e$. A grid spacing of 0.01 $\text{\AA}$ is used for all calculations in order to converge energy eigenvalues to within 0.05 eV of the analytic solutions of the PIB. Though the KS eigenvalues and eigenfunctions can be obtained analytically, we use the values obtained from Octopus for consistency in comparing to the ensemble-generalized LSDA HXC values which we obtain from Octopus.

For each choice of $N_e$, we first run a spin-polarized GS calculation for independent particles in 1D, calculating $N_e/2+1$ states to include all the filled states plus one unoccupied state. %
We then run a \enquote{one-shot} DFT calculation with the same value of $N_e$, but occupations of the KS states for each state in the ensemble are built based on $f^{m}_{j}$ of equation (\ref{eqn:ensdens}), which are obtained from Slater determinants as outlined in section \ref{subsec:Slater}. These calculations use fixed wavefunctions from the previous independent-particles calculation, and provide $E_{\rm H}$, $E_{\rm x}$, and $E_{\rm c}$ for a density built from the given occupations. 

Given the problematic nature of the Coulomb interaction in 1D, we describe the electron-electron interactions with the 1D soft Coulomb potential, where we set the softening parameter, $a$, to 1 Bohr radius ($a_0$):
\begin{eqnarray}
v_\text{sc}(x) = \frac{1}{\sqrt{x^2 + a^2}}.
\end{eqnarray}
We use the 1D LSDA exchange\cite{HFC11} and correlation functionals\cite{CSS06} as implemented in $\texttt{libxc}$  4.3.4, \cite{libxc} which were parametrized for this interaction and value of $a$.

\subsection{\label{sec:Bi_e}Bi-ensemble: Symmetry-enforced}

Starting from equation (\ref{eqn:Hxc}), the GOK weighting scheme from equation (\ref{eqn:GOKweights}), and the multiplet structure of figure \ref{fig:mult} with a choice of the bi-ensemble ($I=2$, $g_2=3$ and $M_2=4$), our weights are:
\begin{equation}
\mathtt{w}_m =
    \begin{cases}
            1-3\mathtt{w} & m \leq 1,  \\
            \mathtt{w} &   m > 1.
    \end{cases}
    \label{eqn:bi_weights}
\end{equation}
The corresponding excitation energy correction from equation (\ref{eqn:general_correction}), as in equation (\ref{eqn:correctionline}) for the UEG, is 
\begin{equation}
\left. \frac{\partial E_{\text{Hxc}}^{w, I=2} [\rho]}{\partial \mathtt{w}} \right|_{\rho=\rho^w_{I=1}} = -3 E_{\text{Hxc}}[\rho_{\text{GS}}] + E_{\text{Hxc}}[\rho_{\alpha \alpha}] \\ + E_{\text{Hxc}}[\rho_{\beta \beta}] + E_{\text{Hxc}}[\rho_{\alpha \beta + \beta \alpha}] .
\label{eqn:bi_e_expanded}
\end{equation}
To use this expression directly would break the spin-symmetry of the triplet, as noted in Section \ref{subsec:Slater}. We note that other EDFT methods have avoided this symmetry-breaking issue via approximations based on multi-determinant spin eigenstates rather than just the density.\cite{GPG02, PYTB14} To enforce spin symmetry, we use the energy $E_{\text{Hxc}}[\rho_{\alpha \alpha}]$ for all states in the triplet, simplifying equation \ref{eqn:bi_e_expanded} to:
\begin{equation}
\frac{\partial E_\text{Hxc}^{w,I=2}[\rho]}{\partial \mathtt{w}}  \bigg\rvert_{\rho=\rho^w_{I=2}} = -3 E_{\text{Hxc}}[\rho_{\text{GS}}] + 3 E_{\text{Hxc}}[\rho_{\alpha \alpha}],
\label{eqn:bicorrectionsimplified}
\end{equation}
as was written similarly in equation (\ref{eqn:correctionline}), where the coefficient of 3 on the second term reflects the degeneracy of the highest multiplet, which we have enforced in this section. The difference of KS energies $E_2 - E_1$ can be reduced to a difference of eigenvalues \textit{via} equation (\ref{eqn:EmthKS}):
\begin{equation}
    E_2 - E_1 = \sum^\infty_{j=1} f^{2}_{j} \epsilon_j - \sum^\infty_{j=1} f^{1}_{j} \epsilon_j .
    \label{eqn:engdiff}
\end{equation}
This expression reduces to the same result for both the bi-ensemble and, as needed later in section \ref{sec:Tri_pres}, the tri-ensemble -- that is, $E_2 - E_1 = E_3 - E_1 = \epsilon_{n=N_e/2+1} - \epsilon_{n=N_e/2}$. 
The first excitation energy from equation (\ref{eqn:omega_1}), denoted $\Omega_2^{\text{e}}$ with `e' for symmetry-enforced approach, is:
\begin{equation}
\Omega_{1}^{\text{e}} = \epsilon_{n=N_e/2+1} - \epsilon_{n=N_e/2} -  E_{\text{Hxc}}[\rho_{\text{GS}}] + E_{\text{Hxc}}[\rho_{\alpha \alpha}].
\label{eqn:Omega1e}
\end{equation}
\begin{figure*}[ht]	
\centering{
\begin{tikzpicture}
\node [anchor=north west] (imgA) at (-0.10\linewidth,.58\linewidth){\includegraphics[width=0.49\linewidth]{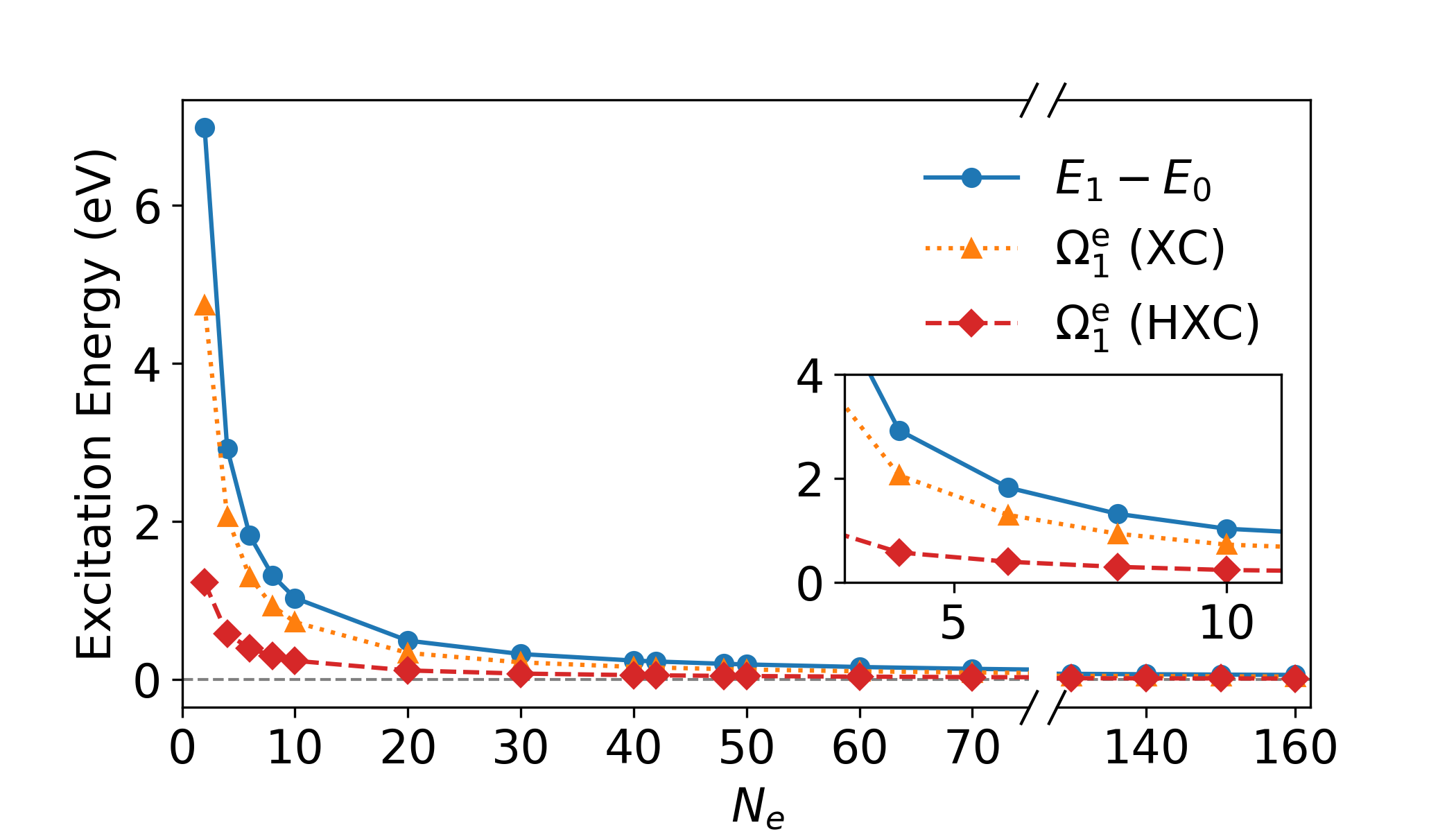}};
\node [anchor=north west] (imgB) at (0.40\linewidth,.58\linewidth){\includegraphics[width=0.49\linewidth]{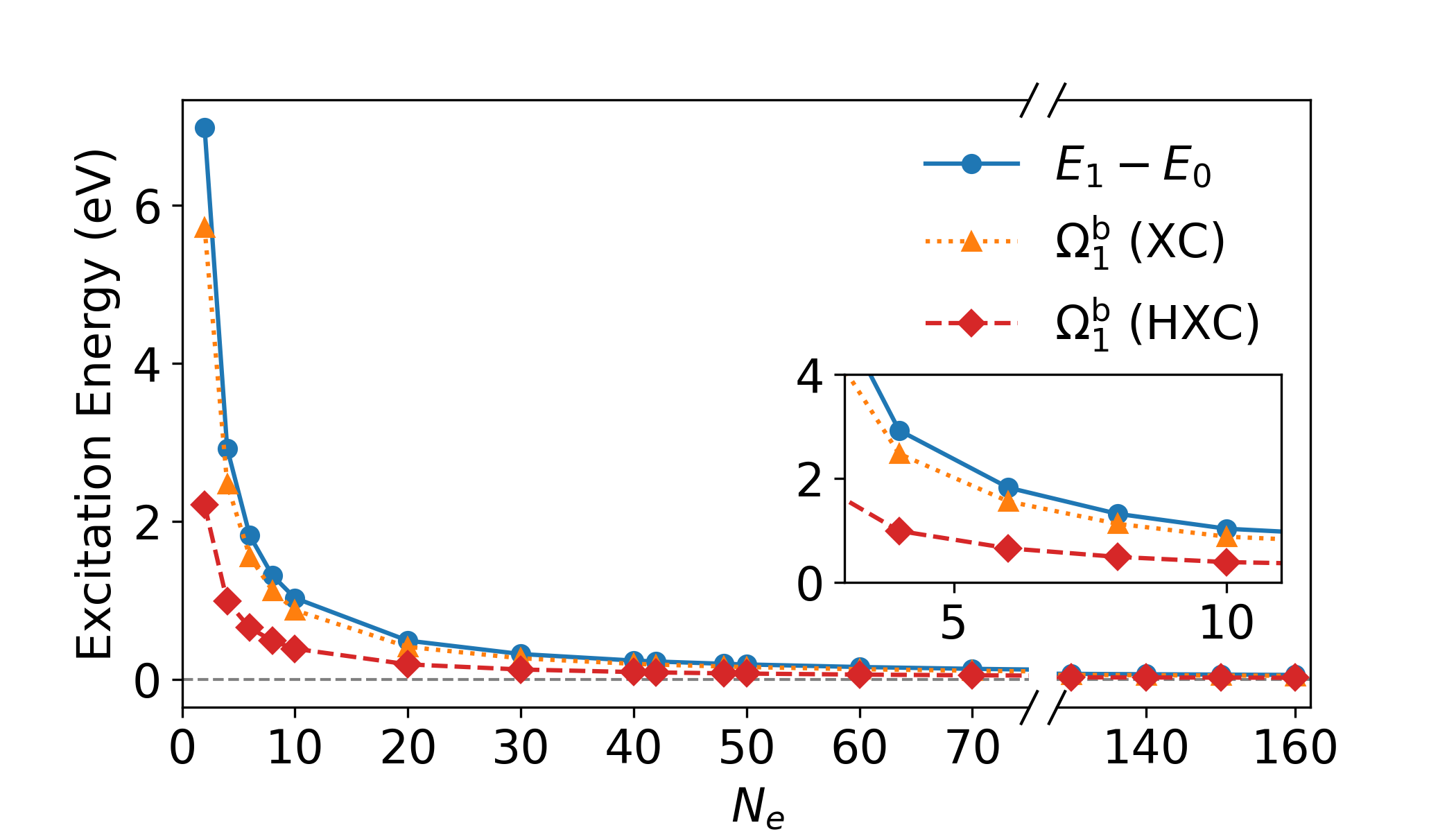}};    
\draw [anchor=north west] (-0.10\linewidth, .58\linewidth) node {\textbf{(a)} Symmetry-enforced bi-ensemble};
\draw [anchor=north west] (0.40\linewidth, .58\linewidth) node {\textbf{(b)} Symmetry-broken bi-ensemble};
\end{tikzpicture}}\vspace{-0.5cm}\caption{Ensemble-corrected first excitation energies compared to KS energy differences for: \textbf{(a)} the triplet $\Omega_1^\text{e}$ of the symmetry-enforced bi-ensemble described in section \ref{sec:Bi_e} and by equation (\ref{eqn:Omega1e}); \textbf{(b)} the triplet $\Omega_1^\text{b}$ of the symmetry-broken bi-ensemble described in section \ref{sec:Bi_b} and by equation (\ref{eqn:Omega1b}). The labels `e' and `b' denote results from the symmetry-enforced and symmetry-broken ensembles. HXC denotes results with a weight-dependent Hartree, while XC denotes the use of a \enquote{traditional} Hartree, as explained in section \ref{subsec:HXC}.}
 		\label{fig:bi}			
 \end{figure*}
\subsection{\label{sec:Bi_b}Bi-ensemble: Symmetry-broken}
In a second alternative method, we do not enforce any symmetry, and only simplify equation (\ref{eqn:tricorrection}) based on equalities that are satisfied in practice by LSDA. We use $E_{\text{Hxc}}[\rho_{\alpha \alpha}]$ for only two states in the triplet. $E_{\text{Hxc}}[\rho_{\alpha \beta \pm \beta \alpha}]$ is then used for the third state of the triplet, breaking the spin symmetry. We use the same weights as in equation (\ref{eqn:bi_weights}) and obtain:
\begin{equation}
\frac{\partial E_\text{Hxc}^{w,I=2}[\rho]}{\partial \mathtt{w}}  \bigg\rvert_{\rho=\rho^w_{I=2}} = -3 E_{\text{Hxc}}[\rho_{\text{GS}}] + 2 E_{\text{Hxc}}[\rho_{\alpha \alpha}] + \\ E_{\text{Hxc}}[\rho_{\alpha\beta \pm \beta\alpha}].
\label{eqn:bicorrectionbroken}
\end{equation}
The first excitation energy from equation (\ref{eqn:omega_1}), denoted $\Omega_1^{\text{b}}$ with `b' for symmetry-broken approach, is then calculated as:
\begin{equation}
\Omega_{1}^{\text{b}} = \epsilon_{n=N_e/2+1} - \epsilon_{n=N_e/2} - E_{\text{Hxc}}[\rho_{\text{GS}}] + \frac{2}{3} E_{\text{Hxc}}[\rho_{\alpha \alpha}] + \frac{1}{3}E_{\text{Hxc}}[\rho_{\alpha\beta \pm \beta\alpha}]
\label{eqn:Omega1b}
\end{equation}
The difference between triplet energies from the symmetry-enforced and symmetry-broken bi-ensembles, from equations (\ref{eqn:Omega1e}) and (\ref{eqn:Omega1b}) is:
\begin{equation}
\Omega_1^{\text{e}} - \Omega_1^{\text{b}} =     \frac{1}{3} E_{\text{xc}}[\rho_{\alpha \alpha}] - \frac{1}{3}E_{\text{xc}}[\rho_{\alpha\beta \pm \beta\alpha}] .
\label{eqn:bi_Omega_diff}
\end{equation}
Because the Hartree term is spin-independent, its value is the same when evaluated on $\rho_{\alpha\alpha}$ and $\rho_{\alpha\beta \pm \beta\alpha}$. For this reason, the difference in corrected excitation energies obtained in equation (\ref{eqn:bi_Omega_diff}) only has a contribution from XC, and is the same whether an ensemble-generalized Hartree is used or not.
\subsection{\label{sec:Tri_pres}Tri-ensemble}
We now consider a tri-ensemble, $I=3$, based on figure \ref{fig:mult}. In order to calculate the singlet energy, $\Omega_2$, we begin with equation (\ref{eqn:final_omega_2}). Knowing the difference of non-interacting energies from the PIB,
all that is left is to calculate the derivative of $E_\text{Hxc}$.
Given the multiplet structure of figure \ref{fig:mult}A with $g_3=1$ and $M_3=5$, we have weights
\begin{equation}
\mathtt{w}_m =
    \begin{cases}
            \frac{1-\mathtt{w}}{4} & m \leq 4,   \\
            \mathtt{w} &   m > 4.
    \end{cases}
    \label{eqn:weights2}
\end{equation}
The $I=3$ derivative with respect to the weight is:
\begin{equation}
    \left. \frac{\partial E_{\text{Hxc}}^{w, I=3} [\rho]}{\partial \mathtt{w}} \right|_{\rho=\rho^w_{I=3}} = -\frac{1}{4} ( E_{\text{Hxc}}[\rho_{\text{GS}}] + E_{\text{Hxc}}[\rho_{\alpha \alpha}] \\ + E_{\text{Hxc}}[\rho_{\beta \beta}] + E_{\text{Hxc}}[\rho_{\alpha\beta + \beta\alpha}] ) + E_{\text{Hxc}}[\rho_{\alpha\beta - \beta\alpha}].
\label{eqn:tricorrection}
\end{equation}
\hide{Method 2 is again another one where we use an $\mathtt{Occupations}$ block of all 0.5's, corresponding to equation (\ref{eqn:Ms0dens}):}
As done for the bi-ensemble in section \ref{sec:Bi_e}, we again enforce spin symmetry by using the energy $E_{\text{Hxc}}[\rho_{\alpha \alpha}]$ for all states in the triplet.
The last term, representing the singlet, we write as $E_{\text{Hxc}}[\rho_{\alpha \beta \pm \beta \alpha}]$.
The second excitation energy (i.e. the singlet), %
then is calculated by combining equations (\ref{eqn:final_omega_2}), (\ref{eqn:tricorrection}), and (\ref{eqn:bicorrectionsimplified}) from the symmetry-enforced bi-ensemble:
\begin{equation}
    \Omega^{\rm}_2 \approx \epsilon_{n=N_e/2+1} - \epsilon_{n=N_e/2} - E_{\text{Hxc}}[\rho_{\text{GS}}] + E_{\text{Hxc}}[\rho_{\alpha\beta \pm \beta\alpha}].
    \label{eqn:Omega_e}
\end{equation}
\begin{table*}
\caption{\label{tab:widetable}Ensemble-corrected excitation energies tabulated versus number of electrons $N_e$, compared with KS energy difference $E_2-E_1 = E_3-E_1$, all reported in eV. $\Omega_1$ is the first excitation energy (triplet), obtained from the bi-ensemble. $\Omega_{2}$ is the second excitation energy (singlet), obtained from the tri-ensemble. The labels `e' and `b' denote results from the symmetry-enforced and symmetry-broken ensembles. HXC denotes results with a weight-dependent Hartree, while XC denotes the use of a \enquote{traditional} Hartree, as explained in section \ref{subsec:HXC}.}
\begin{ruledtabular}
\footnotesize
\begin{tabular}{rr|rr|rr|rr}&\multicolumn{1}{c|}{}
 &\multicolumn{2}{c|}{$\Omega_1^{\text{e}}$}&\multicolumn{2}{c|}{$\Omega_1^{\text{b}}$}&\multicolumn{2}{c}{$\Omega^{\text{}}_2$}\\ \hline
 $N_e$&$E_2 - E_1$&XC&HXC&XC&HXC&XC&HXC\\ \hline
 2  &6.988  &4.741 &1.229 &5.722  & 2.210 & 7.684   & 4.172 \\
 4  &2.925  &2.061 &0.5767&2.474  & 0.9901& 3.301   & 1.817 \\
 6  &1.822  &1.298 &0.3970&1.557  & 0.6563& 2.076   & 1.175 \\
 8  &1.319  &0.9356&0.2989&1.127  & 0.4902& 1.509   & 0.8727\\
 10 &1.032  &0.7268&0.2382&0.8793 & 0.3906& 1.184   & 0.6954\\
 20 &0.4931 &0.3372&0.1159&0.4140 & 0.1928& 0.5676  & 0.3464\\
 30 &0.3236 &0.2179&0.0761&0.2695 & 0.1277& 0.3727  & 0.2309\\
 40 &0.2408 &0.1607&0.0566&0.1996 & 0.0954& 0.2774  & 0.1732\\
 42 &0.2291 &0.1527&0.0538&0.1897 & 0.0909& 0.2639  & 0.1650\\
 48 &0.1999 &0.1327&0.0469&0.1652 & 0.0794& 0.2302  & 0.1444\\
 50 &0.1917 &0.1272&0.0450&0.1584 & 0.0762& 0.2208  & 0.1386\\
 60 &0.1592 &0.1052&0.0373&0.1313 & 0.0634& 0.1834  & 0.1155\\
 70 &0.1362 &0.0897&0.0319&0.1121 & 0.0543& 0.1568  & 0.0990\\
 80 &0.1189 &0.0782&0.0279&0.0978& 0.0475& 0.1370  & 0.0867\\
 90 &0.1056 &0.0693&0.0248&0.0867& 0.0422& 0.1216  & 0.0770\\
 100&0.0949 &0.0622&0.0222&0.0779 & 0.0379& 0.1093  & 0.0693\\
 110&0.0862 &0.0564&0.0202&0.0707 & 0.0345& 0.0993  & 0.0631\\
 120&0.0780 &0.0516&0.0185&0.0647 & 0.0316& 0.0910  & 0.0578\\
 130&0.0729 &0.0476&0.0171&0.0597 & 0.0292& 0.0839  & 0.0534\\
 140&0.0676 &0.0441&0.0158&0.0554 & 0.0270& 0.0779  & 0.0496\\
 150&0.0631 &0.0412&0.0148&0.0517 & 0.0253& 0.0726  & 0.0463\\
 160&0.0591 &0.0385&0.0139&0.0484 & 0.0237& 0.0681  & 0.0434\\
\end{tabular}
\end{ruledtabular}
\end{table*}
Regardless of whether the symmetry-broken or symmetry-enforced approach is used, the same result is obtained. %
\begin{figure}
\includegraphics[width=0.8\linewidth]{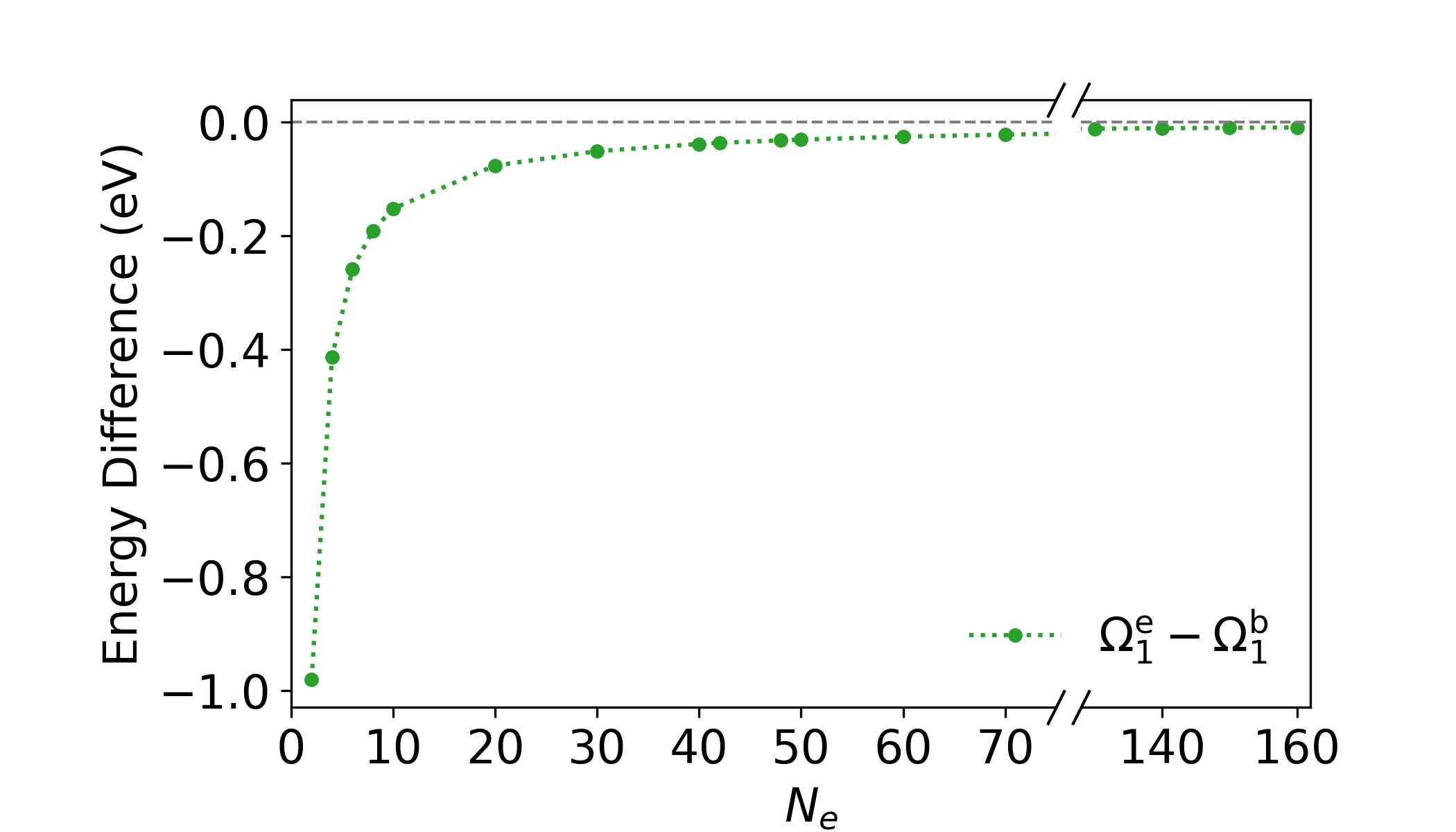}
\caption{Difference between excitation energies of  triplet excited states from the symmetry-enforced ($\Omega^e_1$) and symmetry-broken ($\Omega^b_1$) bi-ensembles, as given in equation (\ref{eqn:bi_Omega_diff}). The labels `e' and `b' denote results from the symmetry-enforced and symmetry-broken ensembles.}
 		\label{fig:tri_diff}		
 \end{figure}
\begin{figure}
\includegraphics[width=120mm,scale=0.5]{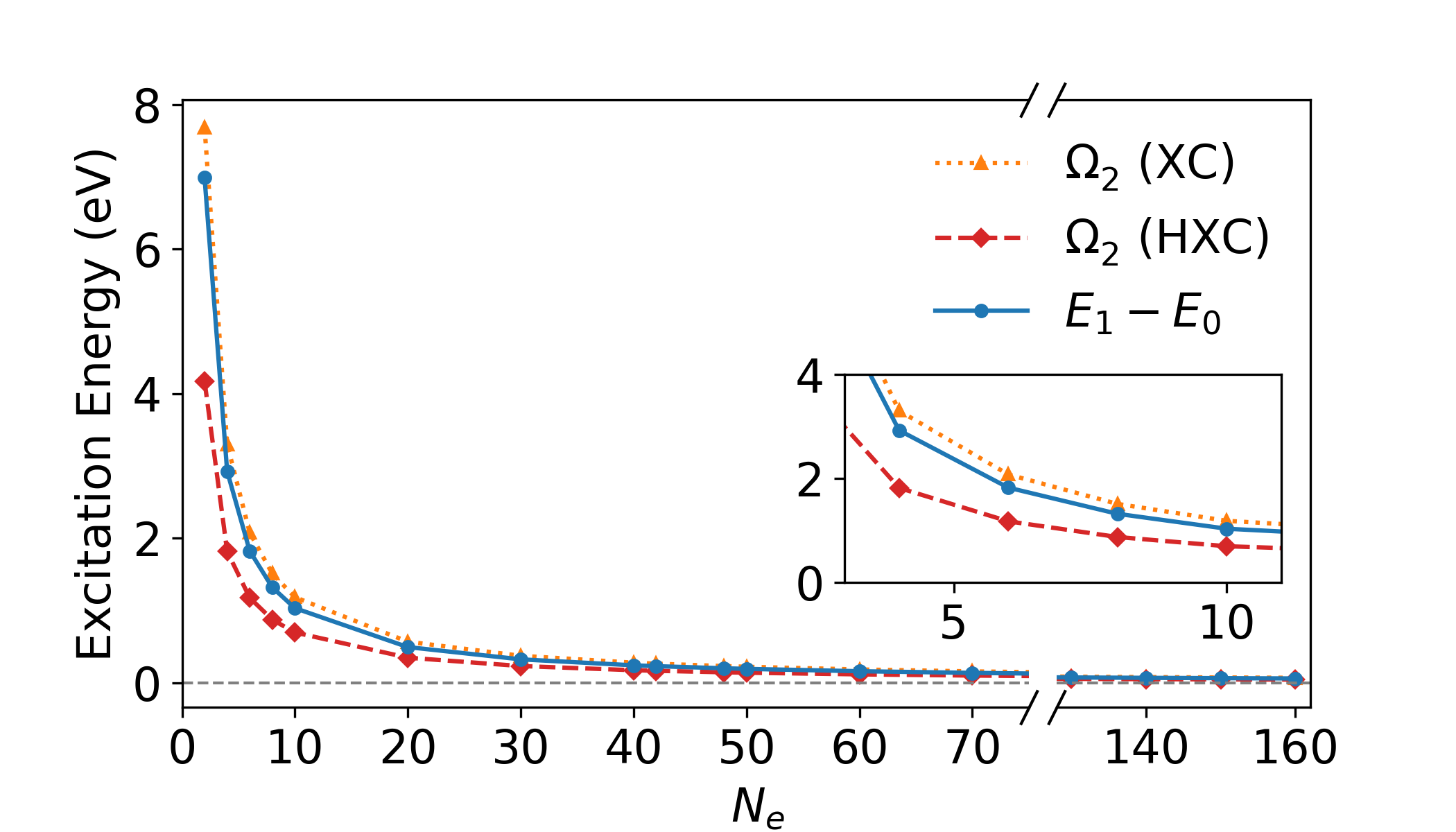}%
\vspace{-0.40cm}\caption{Energy of singlet states obtained as second excitation energies $\Omega_2$ from the tri-ensemble described by equation (\ref{eqn:Omega_e}), compared to KS energy differences $E_2 \minus E_1$. Insets show detail of regions for small $N_e$. HXC denotes results with a weight-dependent Hartree, while XC denotes the use of a \enquote{traditional} Hartree, as explained in section \ref{subsec:HXC}.}
 		\label{fig:tri}		
\end{figure}
The second excitation energy can also be computed by means of a symmetry-projected ensemble (as in references \citenum{PYTB14} and \citenum{AJC}), comprising of states with the same spin-symmetry only. The equation for the ensemble-corrected singlet energy as obtained from the symmetry-projected ensemble can be obtained using equation (34) of reference \citenum{GOK88I} for non-degenerate ensembles. Equation (\ref{eqn:Omega_e}) for the singlet energy is once again obtained from this approach, showing the consistency of these three ensemble approaches.

\section{\label{sec:Results}Results and Discussion}
\subsection{\label{sec:bi_inst_results}Triplet Energies from Bi-ensembles}

The corrected first excitation energies $\Omega_1$ from the bi-ensemble, according to the symmetry-enforced and symmetry-broken schemes, are shown in figure \ref{fig:bi}. All numerical results are also tabulated in Table \ref{tab:widetable}. We find that in these cases, and all cases we study in this paper, the excitation energies go to zero in the thermodynamic limit, in agreement with the metallic expectation from Luttinger liquid theory.\cite{Haldane_1981,Imambekov}
Triplet energies calculated from both the symmetry-broken and symmetry-enforced bi-ensembles are smaller than the non-interacting energy difference but are positive for all $N_e$. When a \enquote{traditional} Hartree DFA is used (the XC case), larger excitation energies are obtained than in the ensemble-generalized LSDA HXC case. The energy difference between symmetry-enforced and symmetry-broken is plotted in figure \ref{fig:tri_diff}. 

We find positive excitation energies in all cases in this paper, indicating a lack of a triplet instability, consistent with the known singlet ground state.\cite{Lieb-Mattis} This is a point in favor of EDFT since triplet instabilities are known to exist in other theories, such as Hartree-Fock,\cite{PhysRevLett.4.415,10.1143/PTP.25.653,https://doi.org/10.1002/qua.560010413} time-dependent Hartree-Fock (TDHF),\cite{doi:10.1021/cr0505627, 10.1063/1.1701562} and TDDFT,\cite{Peach} and triplet instabilities have also been reported in the 3D electron gas at metallic densities.\cite{PhysRevB.3.1910} 
\subsection{Singlet Energies from Tri-ensembles}
All ensemble-corrected singlet excitation energies from the tri-ensemble are positive (figure \ref{fig:tri}). The corrected second excitation energy $\Omega_2^{\text{}}$ is lower in value than the KS second excitation energy, computed as $E_2 \minus E_1$, when weight-dependent Hartree (HXC) is used, and greater than the KS energy difference when \enquote{traditional} Hartree DFA is used (XC). %
\subsection{\label{sec:effmass} Effective Masses}
\begin{figure*}[ht]	
\centering{
\begin{tikzpicture}
\node [anchor=north west] (imgA) at (-0.10\linewidth,.58\linewidth){\includegraphics[width=0.48\linewidth]{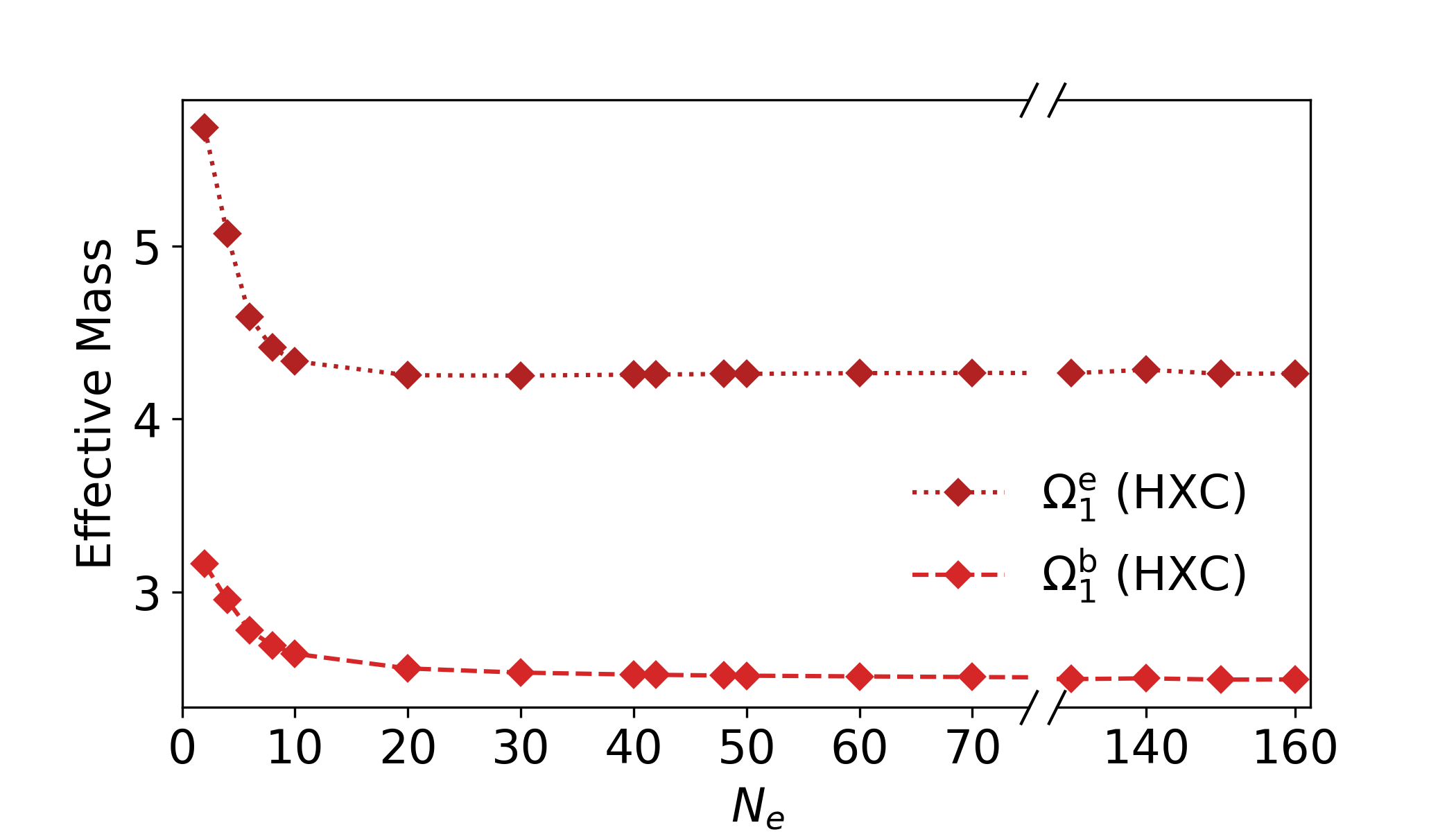}};
\node [anchor=north west] (imgB) at (0.40\linewidth,.58\linewidth){\includegraphics[width=0.48\linewidth]{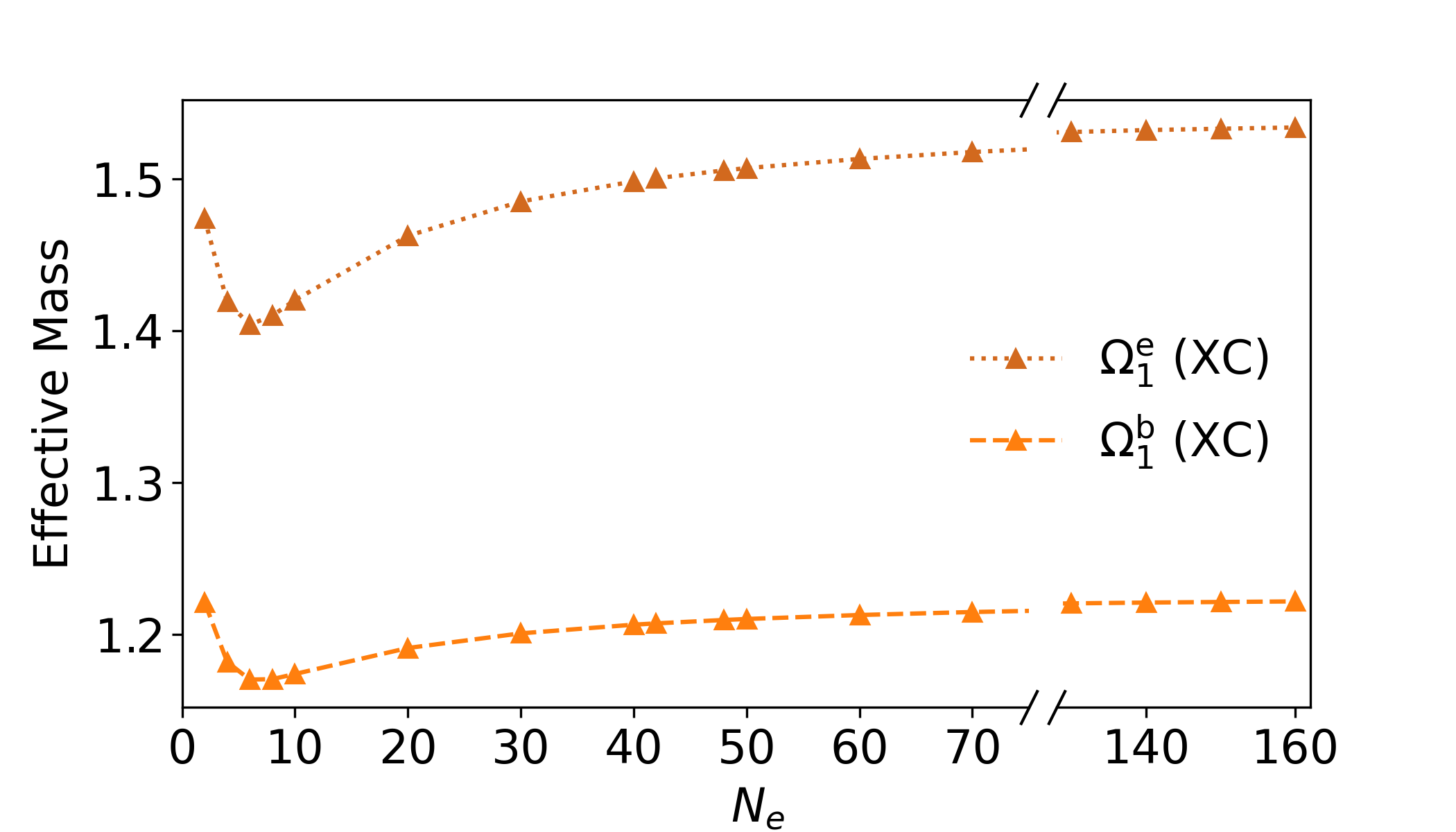}};
\draw [anchor=north west] (-0.10\linewidth, .58\linewidth) node {\textbf{(a)} Bi-ensemble, weight-dependent Hartree};
\draw [anchor=north west] (0.41\linewidth, .58\linewidth) node {\textbf{(b)} 
Bi-ensemble, \enquote{Traditional} Hartree};
             \end{tikzpicture}
 			}
 	\caption{Effective masses, calculated from equation (\ref{eqn:mass}), for the symmetry-enforced bi-ensemble described in section \ref{sec:Bi_e} and the symmetry-broken bi-ensemble described in section \ref{sec:Bi_b} in the \textbf{(a)} HXC, and \textbf{(b)} XC cases. The labels `e' and `b' denote results from the symmetry-enforced and symmetry-broken ensembles. HXC denotes results with a weight-dependent Hartree, while XC denotes the use of a \enquote{traditional} Hartree, as explained in section \ref{subsec:HXC}.}
 		\label{fig:mass}	
 \end{figure*}
\begin{figure}
\includegraphics[width=0.7\linewidth]{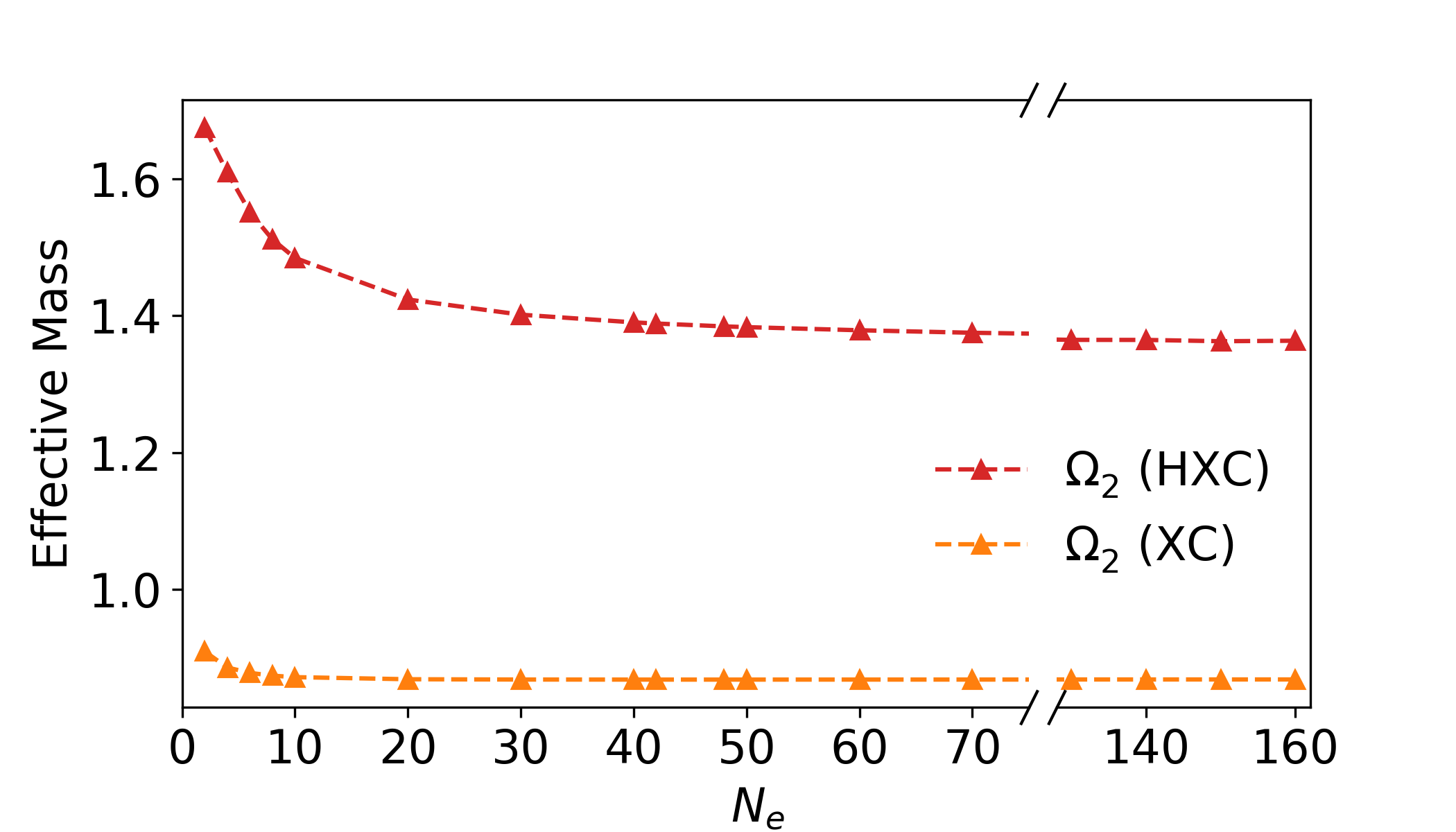}
\caption{Effective masses, calculated from equation (\ref{eqn:mass}), for the tri-ensemble. HXC denotes results with a weight-dependent Hartree, while XC denotes the use of a \enquote{traditional} Hartree, as explained in section \ref{subsec:HXC}.}
\label{fig:trimass}
\end{figure}
Given that our excitation energies go to zero in the thermodynamic limit, we cannot meaningfully study corrections to the bandgap, but we instead investigate the effective mass to look for non-vanishing corrections.
The effective mass is a useful parameter by which to validate a model's treatment of interactions, and it can be directly studied in real systems. For instance, studies with different theories have tried to reproduce the experimentally measured occupied bandwidth of sodium, with varying success.\cite{PhysRevB.68.165103,northrup} 
The first and second excitation energies become identical in the metallic continuum limit (i.e. no singlet-triplet splitting) which is why we treat all effective masses as estimates of the same quantity, and can compare our results approaching the thermodynamic limit to the UEG. 
To compute an effective mass in our case, we consider the excitation as an energy difference between $k$-points $k_2$ and $k_1$ on parabolic bands in a UEG, as in equation (\ref{eqn:kEnergy}). We assume that the $k$-points for the excitation are the same for the non-interacting system and the interacting, ensemble-corrected system, which is appropriate if the Fermi level is not shifted with respect to the states by the interaction, like the condition for a conserving approximation in many-body perturbation theory.\cite{Baym}  
With these considerations, we obtain the effective mass as a ratio between the independent-particle and interacting excitation energies:
\begin{equation}
\frac{\Delta E^{\text{ip}}}{\Delta E^{\text{int}}} = \frac{E_2 - E_1}{\Omega} = \frac{\frac{\hbar^2 {k_2}^2}{2m_e} \minus \frac{\hbar^2 {k_1}^2}{2m_e}}{\frac{\hbar^2 {k_2}^2}{2m^\text{*}} \minus \frac{\hbar^2 {k_1}^2}{2m^\text{*}}} = \frac{m^{\text{*}}}{m_e}.
\label{eqn:mass}
\end{equation}

While formally the effective mass is only defined for periodic systems, we study the behavior of this ratio for our finite systems  and the limit as our model approaches a periodic system, which may be compared to the use of bulk effective masses in studying quantum dots.\cite{Efros}
The effective masses at the thermodynamic limit (estimated as the results at our largest $N_e$, 160) are reported in Table \ref{tab:table}. In each case, the electron mass ratio approaches a limit different from 1.
By contrast, it can be shown analytically that GS DFT with LDA gives the effective mass in a UEG always equal to the free electron mass.\cite{LOUIE20069} That a nontrivial change in the effective mass is found through EDFT shows the promise of EDFT for periodic systems, and the promise for additional insight to be obtained through the use of more sophisticated ensemble DFAs. 

All effective masses obtained are positive in value, indicating electron, rather than hole, character which is expected for a metallic system. In the XC case for the singlet, we find an effective mass of 0.8684, notably <$1$, while the results obtained from all other cases are >$1$.
The effective masses for the bi-ensemble exhibit several different behaviors (figure \ref{fig:mass}). For both the symmetry-enforced and symmetry-broken bi-ensembles with ``traditional'' Hartree (XC), the effective masses are >1 and decrease from their value at $N_e = 2$ to a minimum at $N_e = 6$, after which point the effective mass increases slightly as it converges to its limit. By contrast, HXC values decrease monotonically from $N_e$ = 2.
We find positive and monotonically decreasing effective masses for the tri-ensemble. The use of weight-dependent Hartree in the tri-ensemble results increases the effective mass by a fairly constant value of 0.5, as shown in figure \ref{fig:trimass}. For both bi-ensemble and tri-ensemble, the HXC results are systematically larger than for XC, and the symmetry-enforced results are systematically larger than the symmetry-broken results.
We can identify $E_{\rm xc} \left[ \rho_{\alpha \alpha} \right]$ as the cause of the non-monotonic behavior in the bi-ensemble with ``traditional'' Hartree: this term has a sharp decrease in magnitude (becoming less negative) at low $N_e$, faster than $N_e$, but it is outweighed by the Hartree terms in the weight-dependent Hartree case, and it is absent in the tri-ensemble case. It is not clear what deeper meaning might be associated with this non-monotonic behavior.

We are not aware of any reported values for the effective mass of electrons in the 1D UEG with soft Coulomb interactions. Instead we have two points of comparison. First, for a 1D UEG with a contact interaction $V(x-x') = V_0 \delta \left( x - x' \right)$, a GW calculation\cite{Tanatar} found $m^*$ in the range 1 to 2.5 (and presumably continuing to increase), depending on $V_0$ and the density. These values are comparable to most of our results, ranging between 1 and 5. Second, there is extensive literature for 2D and 3D UEGs. Conventionally, UEGs are characterized by the density parameter $r_s$, which is the Wigner-Seitz radius measured in Bohr radii $a_0$. The 1D generalization \cite{CSS06} is $r_s = a_0/2 \rho$, which in our case is 1.89. We compared our results to the effective mass for the UEG obtained \textit{via} Monte Carlo for 2D and 3D systems with $r_s \leq 4$, representing the metallic regime, where $r_s \leq 1$ represents the high-density regime.\cite{Haule, ParrYang} In the 3D case, effective masses in the UEG obtained by variational diagrammatic Monte Carlo (MC) have been found to be 0.955(1) for $r_s = 1$, and 0.996(3) at $r_s = 4$.\cite{Haule} Other calculations on the 3D UEG done via diffusion MC extrapolated to the thermodynamic limit have reported an effective mass of 0.85 at $r_s \approx 4$.\cite{PhysRevLett.127.086401} For a 2D UEG, diffusion MC gave results for a paramagnetic case of 0.955(2) at $r_s = 1$ and 1.04(2) at $r_s = 5$.\cite{PhysRevB.87.045131} The ferromagnetic case gave 0.851(5) at $r_s = 1$ and 0.74(1) at $r_s = 5$;\cite{PhysRevB.87.045131}. In the high-density limit for a 3D electron gas, the effective mass is expected to be less than one.\cite{PhysRevB.53.7352} Our 1D result of 0.8684 from the tri-ensemble with weight-independent Hartree (XC) is fairly similar to the 2D and 3D cases, which seems reasonable given the weak dependence on dimensionality seen between 2D and 3D, and the spread in literature values for the effective masses. More conclusive assessment of the accuracy of our effective masses would need a reliable calculation with another method for the 1D UEG with soft-Coulomb interactions.
\bgroup
\def\arraystretch{1.5}
\begin{table}
\caption{\label{tab:table}Effective masses in the thermodynamic limit, estimated from $N_e = 160$, obtained from equation (\ref{eqn:mass}) using the ensemble-corrected excitation energies. The labels `e' and `b' denote results from the symmetry-enforced and symmetry-broken ensembles. HXC denotes results with a weight-dependent Hartree, while XC denotes the use of a \enquote{traditional} Hartree, as explained in section \ref{subsec:HXC}. %
} %
\begin{ruledtabular}
\begin{tabular}{lrr}
Excitation%
&\multicolumn{1}{c}{HXC}%
&\multicolumn{1}{c}{XC}\\
\hline
$\Omega_1^\text{e}$ (singlet)  &  4.263 &  1.534 \\
$\Omega_1^\text{b}$ (singlet)  &  2.495 & 1.222 \\
$\Omega_2$ (triplet) & 1.363 & 0.8684 \\
\end{tabular}
\end{ruledtabular}
\end{table}
\egroup
\section{\label{sec:Conclusion}Conclusion}

Since EDFT was designed for the treatment of discrete energy levels, it does not readily adapt to the band structure of solids. We have therefore instead approached the application of EDFT to a periodic system through a set of systems having the same fixed average density, and studied its approach to the thermodynamic limit. We have considered ensemble-corrected excitation energies for systems where the KS potential is set to the PIB potential, becoming the UEG in the thermodynamic limit, and avoiding the need for SCF calculations.

Corrections to the singlet energy obtained from a tri-ensemble are positive in the XC case, increasing the KS energy differences, and negative in the HXC case. In both the symmetry-enforced bi-ensemble, with \enquote{traditional} Hartree and the symmetry-broken bi-ensemble with \enquote{traditional} Hartree, corrections are smaller than those obtained with weight-dependent Hartree, both results decreasing the KS energy difference. While EDFT provides nonzero corrections to excitation energies in the finite regime, in the approach to the thermodynamic limit, these tend to zero, as do the KS energy differences as well, which is expected for a metallic system.\cite{Kaxiras} We consider symmetry-enforced and symmetry-broken schemes of handling the triplet states that are indistinguishable in density, and find that for the bi-ensemble the symmetry-enforced case leads to larger corrections to the KS energy difference. The excitation energies are positive in all cases, showing no sign of the triplet instability that can show up in some theories.

Effective masses for each of the methods were calculated, and found to approach a positive limit in all cases. A non-trivial correction to the effective mass is found in the thermodynamic limit, even with our simple Hartree and LSDA XC approximations. These results indicate the potential of EDFT in the periodic limit to provide meaningful results. %

Prior work by Kraisler and Kronik\cite{KK14} examined the  derivative discontinuity of XC functionals (which corrects the KS gap) in the thermodynamic limit, based on ensemble considerations (but not on GOK EDFT).
They note that the Hartree-based contribution to the missing derivative discontinuity vanishes in the thermodynamic limit,  with the exact XC component being the source of a useful correction. They note that, as we see in our results, LDA-based corrections to the gap vanish due to known insufficiencies in this approximation. By contrast, our work, investigating effective masses as well, found that there can be a nontrivial correction from LDA in the thermodynamic limit. 

We have investigated the impact of using two different forms of ensemble-generalized Hartree, one in which there is explicit weight dependence in the functional, which is then applied to densities of individual states, and one in which the weight-dependence is only accounted for in the ensemble density (the \enquote{traditional} Hartree). However, it is known that neither method treating the ensemble Hartree is sufficient to treat systems with \enquote{difficult} spin multiplets in finite systems.\cite{GSP20} The weight-dependent Hartree contribution has a significant impact for the singlet energy, changing the sign of the correction. In all cases we have used the former explicitly weight-dependent ensemble-generalized LDA XC. Given that neither LDA nor GGA in periodic systems exhibit the necessary divergence of $f_{\rm xc}$,\cite{ORR02} it is reasonable to expect that implementation of more sophisticated ensemble DFAs, particularly non-local and GIE-free exchange\cite{CPG22} and/or correlation, would be needed for fuller analysis of EDFT's applicability and limitations in treating periodic systems.

While the treatment of increasingly large finite systems at a fixed average density may not be practical for extracting information about real systems, our results from this approach suggest that a formulation of EDFT for periodic systems could provide non-trivial results even with simple DFAs, and motivate further work on finding a suitable formulation. Further study of UEG systems in the thermodynamic limit can be extended to 2D and 3D, with a more realistic Coulomb interaction, as well as to %
models with a nonuniform potential, such as the Kronig-Penney model,\cite{KP} which is not metallic and can be used to investigate whether non-trivial bandgap corrections can be found. Though a lack of density variations for same-spin-symmetry states in the thermodynamic limit presents a challenge for investigating density-driven correlations,\cite{PhysRevLett.123.016401} it may be possible still to extract relevant information from states for which the densities are different in the thermodynamic limit, as in the Kronig-Penney model. Such calculations would require a 
self-consistent EDFT scheme that would accommodate solving for the set of KS potentials needed for matching the set of individual states, as discussed in Ref. \citenum{PhysRevLett.123.016401}.
Finally,
we note that the study of systems with an odd number of electrons, which have a different multiplet structure, may offer further insight into behavior in the thermodynamic limit.

\begin{acknowledgments}
The authors are grateful for the particularly insightful comments and questions of an anonymous referee. R.J.L., D.A.S., and A.P.J. were supported by the U.S. Department of Energy, National Nuclear Security Administration, Minority Serving Institution Partnership Program, under Award DE-NA0003984. R.J.L. was also supported by the NRT program
Convergence of Nano-engineered Devices for Environmental and Sustainable Applications (CONDESA) under NSF award DGE-2125510. Computational resources were provided by the Pinnacles cluster at Cyberinfrastructure and Research Technologies (CIRT), University of California, Merced, supported by the National Science Foundation Award OAC-2019144.
\end{acknowledgments}

\bibliography{aipsamp}%

\end{document}